\documentclass[notitlepage,prd,showkeys,preprintnumbers,floatfix,superscriptaddress,nofootinbib]{revtex4-1}
\usepackage{float}
\usepackage{amsmath,amssymb}
\usepackage{graphicx}
\usepackage{mathtools,slashed}
\usepackage{bm}
\usepackage{comment}
\usepackage{xcolor}
\usepackage{subfigure}
\usepackage{array}
\usepackage{multirow}
\usepackage{url}
\usepackage{color}
\usepackage[T1]{fontenc}
\usepackage[latin9]{inputenc}
\usepackage{graphicx}
\usepackage{esint}
\usepackage{hyperref}
\usepackage{atbegshi}
\usepackage{lipsum}

\begin{document}

\preprint{\vbox{\hbox{CERN-TH-2019-035, CP3-Origins-2019-006 DNRF90
} }}

\title{
Scattering amplitudes from finite-volume spectral functions
} 

\author{John Bulava}
\email[]{bulava@cp3.sdu.dk}
\affiliation{
CP3-Origins, University of Southern Denmark, Campusvej 55, 5230 
Odense M, Denmark
}
\author{Maxwell T. Hansen}
\email[]{maxwell.hansen@cern.ch}
\affiliation{Theoretical Physics Department, CERN, 1211 Geneva 23, Switzerland}

\date{\today}

\begin{abstract}

A novel proposal is outlined to determine scattering amplitudes from finite-volume spectral functions. The method requires extracting smeared spectral functions from finite-volume Euclidean correlation functions, with a particular complex smearing kernel of width $\epsilon$ which implements the standard $i\epsilon$-prescription. In the $L\rightarrow \infty$ limit these smeared spectral functions are therefore equivalent to Minkowskian correlators with a specific time ordering to which a modified LSZ reduction formalism can be applied. The approach is presented for general $m\rightarrow n$ scattering amplitudes (above arbitrary inelastic thresholds) for a single-species real scalar field, although generalization to arbitrary spins and multiple coupled channels is likely straightforward. Processes mediated by the single insertion of an external current are also considered. Numerical determination of the finite-volume smeared spectral function is discussed briefly and the interplay between the finite volume, Euclidean signature, and time-ordered $i\epsilon$-prescription is illustrated perturbatively in a toy example. 

 \end{abstract}


\nopagebreak
\maketitle

\section{Introduction}\label{s:int}

The determination of real-time scattering amplitudes from
Euclidean lattice field theory simulations is challenging.
It was pointed out long ago by Maiani and Testa~\cite{Maiani:1990ca} 
that on-shell scattering amplitudes away from threshold cannot be obtained from the 
asymptotic temporal separation of infinite-volume Euclidean correlation functions.
Calculations of finite-volume energies and matrix elements 
in Monte Carlo simulations, however, employ
$n$-point correlation functions in which all time separations
are taken asymptotically large. 
	The challenge of extracting on-shell scattering information is partly
	 resolved by  
 L\"{u}scher's method, in which finite-volume two-hadron energies are 
 related to elastic scattering amplitudes~\cite{Luscher:1990ux} by a 
 determinant 
 equation. This relation is widely used in numerical simulations and has been extended to 
 non-zero total momenta~\cite{Rummukainen:1995vs,Kim:2005gf,Fu:2011xz}, 
 coupled two-hadron scattering channels~\cite{He:2005ey,Lage:2009zv,Bernard:2010fp,Briceno:2012yi,Hansen:2012tf}, asymmetric 
volumes~\cite{Feng:2004ua}, non-zero spin~\cite{Gockeler:2012yj,
Briceno:2014oea,Morningstar:2017spu}, and amplitudes with
external currents~\cite{Lellouch:2000pv,Lin:2001ek,Meyer:2011um,
Briceno:2012yi,Hansen:2012tf,
Briceno:2014uqa,Briceno:2015csa,Briceno:2015tza,Baroni:2018iau}. The extension of the method to treat three-particle 
amplitudes is currently under development~\cite{Briceno:2012rv,
Polejaeva:2012ut,
Hansen:2014eka,
Meissner:2014dea,
Hansen:2015zga,
Briceno:2017tce,Mai:2017bge,Hammer:2017uqm,Hammer:2017kms,Doring:2018xxx,
Briceno:2018aml,Romero-Lopez:2018rcb,Hansen:2019nir,Blanton:2019igq}.  

The main limitations of this  
finite-volume approach are as follows:
\begin{itemize}
	\item It is limited to energies below three or more particle thresholds. 
		Once  
		the three-particle formalism is fully developed, it will be possible to treat all energies below 
		four-particle thresholds. Such restrictions become 
		increasingly severe for lattice QCD simulations as the light quark masses 
		are lowered to their physical values and below. Application of the 
		three-particle formalism also requires as input the 
		two-to-two amplitude over a continuous range of energies, necessitating a parametrization and 
		extrapolation.

	\item The determinant equations for two-to-two amplitudes are 
		block diagonalized in finite-volume irreps of the relevant total-momentum 
		little group, mixing infinite-volume partial waves. These relations are
		infinite dimensional, and are therefore truncated at some partial wave $\ell_{\max}$. The 
		systematic error due to this truncation must be assessed, but is small 
		near threshold due to the angular-momentum barrier. 
		Conversely, higher partial wave amplitudes are sub-leading 
		and thus difficult to determine. Furthermore, this necessary  
		projection onto partial 
		waves and finite-volume irreps prevents the direct 
		calculation of inclusive processes via the optical theorem.

	\item The determination of individual finite-volume energies and 
		matrix elements 
		becomes cumbersome in large spatial volumes. On one ensemble 
		with  $m_{\pi}L = 6.1$ from a 
		recent lattice QCD 
		calculation of 
		elastic pion-pion scattering~\cite{Andersen:2018mau}, 
		$43$ energies are determined, while another recent 
		$\rho\pi$ scattering calculation~\cite{Woss:2018irj} (performed at heavy 
		quark masses so the $\rho$ is stable) 
		employs 141 energies across two volumes with $m_{\pi} L = 10.3$ and $12.4$.
		The increasingly dense finite-volume spectrum  
		and the simultaneously decreasing magnitude of the  
		desired finite volume effects makes the large volume limit of this 
		formalism intractable.  

	\item The determinant condition for a single finite-volume energy 
		involves amplitudes from all total  
		spins and open scattering channels, so that particular initial and 
		final states are selected only 
	 in special situations. A model parametrization is therefore required to 
		simultaneously extract all relevant amplitudes when fitting the 
		finite-volume energies. 

\end{itemize}

Despite these restrictions, substantial progress has been made in lattice calculations of hadron scattering amplitudes. 
Calculations of the energy dependence of elastic scattering amplitudes
between pseudoscalar mesons have provided valuable determinations of the quark-mass dependence of the low-lying resonances~\cite{Andersen:2018mau,Alexandrou:2017mpi,Bulava:2016mks,Fu:2016itp,Guo:2016zos,Briceno:2016mjc,
Wilson:2015dqa,Feng:2014gba,Dudek:2012xn,Pelissier:2012pi,Aoki:2011yj,
Lang:2011mn,Feng:2010es}. 
These results have sufficient statistical precision to begin addressing
systematics due to the finite volume and lattice spacing. 
Meson-baryon and baryon-baryon calculations~\cite{Andersen:2017una,
Lang:2016hnn,Lang:2012db,Detmold:2015qwf,Berkowitz:2015eaa,Orginos:2015aya} are 
considerably less developed while pioneering calculations which treat 
 multiple coupled two-particle channels~\cite{Briceno:2017qmb,Moir:2016srx,Dudek:2016cru,
Wilson:2014cna,Mohler:2013rwa} have been performed. Calculations of transition amplitudes involving external currents are presented in Refs.~\cite{Andersen:2018mau,Alexandrou:2018jbt,Briceno:2016kkp,
Feng:2014gba} and scattering involving vector bosons is treated in 
Refs.~\cite{Romero-Lopez:2018zyy,Woss:2018irj}.

However, the restrictions listed above have so far prevented the quantitative study of scattering amplitudes in many interesting 
systems. For example, potentially exotic hadrons with hidden heavy-flavor 
quantum numbers couple significantly to multiple two-hadron decay channels~\cite{Olsen:2017bmm,Guo:2017jvc,Lebed:2016hpi,Esposito:2016noz,Chen:2016qju}, 
all of which occur above many inelastic thresholds so that 
the currently-available finite-volume formalism is inapplicable. 
A similar problem exists for lattice QCD studies of 
excited baryon resonances in the 
$1-3{\rm GeV}$ region, for which 
polarization observables and higher partial waves are needed 
phenomenologically~\cite{Proceedings:2018uig}. 
Even low-lying hadronic resonances such as the $N(1440)$ (Roper), $\rho(770)$, 
$K^*(892)$, and $\Delta(1232)$ are located above $n\ge3$ hadron thresholds. Although such effects formally must be treated, they are likely less important for resonances with a weak coupling to $n\ge 3$ hadron states. 

The wealth of existing experimental data on hadron scattering, 
ongoing and future 
experiments, and the ever-increasing physical volume
of lattice QCD simulations\footnote{See recent proceedings of the yearly 
lattice conference for reviews on the current state-of-the-art for lattice QCD 
simulations~\cite{Boyle:2017vhi}. Furthermore, several novel 
algorithms~\cite{Luscher:2017cjh,Ce:2016ajy,Ce:2016idq} potentially enable 
significantly larger volumes than used at present.} 
motivate exploration of alternatives to the finite-volume approach.   
A first step in this direction is Ref.~\cite{Hansen:2017mnd} which outlines a method to obtain single-hadron inclusive transition rates above arbitrary 
inelastic thresholds from Euclidean lattice simulations without
employing the finite volume. Ref.~\cite{Hansen:2017mnd} advocates using spectral 
functions obtained in Euclidean time and finite volume, defined as\footnote{The normalization of the unsmeared spectral function differs from the one used in Ref.~\cite{Hansen:2017mnd} but is more appropriate for the smearing kernels considered in this work. In particular, for the complex kernels we introduce below the real part will continue to satisfy Eq.~\ref{e:sp2}.}  
\begin{align}\label{e:sp1}
	C_{\boldsymbol{p}_1\boldsymbol{p}_1}^{ L, J}(\tau) & =  2 E(\boldsymbol p_1) L^3 \langle \boldsymbol{p}_1 | \hat{J} \, 
	{\rm e}^{-\hat{H}\tau} \, \hat{J}^{\dagger} | \boldsymbol{p}_1 \rangle_{L}  =  
	\int_{0}^{\infty} \frac{dE}{\pi} \, \rho_{\boldsymbol{p}_1\boldsymbol{p}_1}^{ L,J}(E)\, \mathrm{e}^{-E\tau} \,,
	\qquad \\[5pt]
	\rho_{\boldsymbol{p}_1\boldsymbol{p}_1}^{L,{J}}(E) & = 2E(\boldsymbol{p}_1) L^3\sum_{n} \pi\, \delta(E-E^L_{n}) \, |
	 \langle \boldsymbol{p}_1 | \hat{J} | n  \rangle_{L}|^2 \,,
\end{align}
where $E(\boldsymbol{p}_2) \equiv \sqrt{\boldsymbol{p}_2^2 + m^2}$ and $m$ the physical mass. Here the `endcap' state $|\boldsymbol{p}_1 \rangle_{ L}$ is a finite-volume 
one-particle state (with other internal indices suppressed), $\hat{J}$ a local 
external current mediating the transition 
(possibly projected onto definite three-momentum), and $\hat{H}$ the 
Hamiltonian. The sum over $n$ runs over 
all finite-volume Hamiltonian eigenstates, with corresponding energies given by $\{E_n^L\}$.

At first glance $\rho_{\boldsymbol{p}_1\boldsymbol{p}_1}^{L, J}(E)$, a sum over 
$\delta$-functions, is 
qualitatively different from its continuous infinite-volume counterpart. 
Ref.~\cite{Hansen:2017mnd} suggests 
bridging this gap via the Backus-Gilbert approach to inverse 
problems~\cite{BG1,BG2} which yields a smeared spectral function 
\begin{gather}\label{e:sp2}
	\hat{\rho}_{\boldsymbol{p}_1\boldsymbol{p}_1}^{L,J,\epsilon}(E) = \int_{0}^{\infty} \frac{d\omega}{\pi} \, 
	\hat{\delta}_{\epsilon}(E,\omega) \, \rho_{\boldsymbol{p}_1\boldsymbol{p}_1}^{L,J}(\omega) \,, \qquad 
	\int_0^{\infty} \frac{d\omega}{\pi}  \, \hat{\delta}_{\epsilon}(E,\omega) = 1 \,.  
\end{gather}
Eq.~\ref{e:sp2} introduces the smearing kernel 
$\lim_{\epsilon \rightarrow 0^+} \hat{\delta}_{\epsilon}(E,\omega) = \pi \delta(E-\omega)$ with a characteristic width $\epsilon$ that compromises 
between resolution and numerical stability. The infinite-volume spectral function, from which the total transition rate is easily obtained, is 
recovered from the ordered double limit 
$\rho_{\boldsymbol{p}_1\boldsymbol{p}_1}^{J}(E) = \lim_{\epsilon \rightarrow 0^+}  \, \lim_{L \rightarrow \infty} 
\hat{\rho}_{\boldsymbol{p}_1\boldsymbol{p}_1}^{L,J,\epsilon}(E)$. 
This procedure nicely circumvents the Maiani-Testa no-go 
theorem~\cite{Maiani:1990ca} by \emph{not} taking the asymptotic time limit 
$t\rightarrow \infty$ typical in calculations of finite-volume energies 
and matrix elements. Although Ref.~\cite{Hansen:2017mnd} provides tests in a toy
model, an application to zero-temperature lattice data has not yet been published. 

This work describes a generalization of Ref.~\cite{Hansen:2017mnd} in which the Lehmann-Symanzik-Zimmermann (LSZ) reduction formalism is used to
determine arbitrary scattering amplitudes from finite-volume spectral 
functions. Processes mediated by a single local external current are also 
included in addition to purely hadronic processes without such currents. As a first 
example, consider the spectral function from 
Eq.~\ref{e:sp1} with pion endcaps and $\hat J$ set to the temporal component 
of the axial current at definite momentum $\boldsymbol{p}_2$. 
The formalism of Ref.~\cite{Hansen:2017mnd} yields the inclusive rate for the process $\pi(\boldsymbol{p}_1) + A_0(\boldsymbol{p}_2) \rightarrow X$ at 
an arbitrary four-momentum transfer determined by the energy argument $E$.

Since $\hat{A}_0(\boldsymbol{p}_2)$ has the quantum numbers of a single pion of momentum 
$\boldsymbol{p}_2$, the rate for $\pi(\boldsymbol{p}_1) + A_0(\boldsymbol{p}_2) \rightarrow X$ develops a pole when the energy carried by the current coincides 
with the on-shell value $E = E(\boldsymbol{p}_2)$. The residue at this pole then gives the purely hadronic inclusive rate $\pi(\boldsymbol{p}_1) + \pi(\boldsymbol{p}_2) \rightarrow X$. Taking this approach further, Sec.~\ref{s:2to2} demonstrates that  
a different choice of smearing kernel can be employed in Eq.~\ref{e:sp2} (to the 
same underlying spectral function) to implement the  
$i\epsilon$-prescription in the analytic continuation to real scattering energies, thereby defining a quantity that coincides with the exclusive scattering amplitude, rather than the inclusive rate, in the ordered double limit.

In contrast to the total rates considered in Ref.~\cite{Hansen:2017mnd}, 
scattering amplitudes are complex-valued functions with the real and imaginary parts 
related by unitarity constraints. However, the correlator 
$C_{\boldsymbol{p}_1\boldsymbol{p}_1}^{ L, J}(\tau)$ is pure real so that 
the complexity enters only through the specific definition of the 
smearing kernel, which takes the form of a complex-valued pole. An important 
result of this work is that the practitioner can choose the value of 
$i \epsilon$ in the pole, and that this parameter is distributed through all 
unitarity cuts  to ensure   a well-defined $L \to \infty$ limit. 
In direct analogy to Ref.~\cite{Hansen:2017mnd}, one requires a hierarchy of
scales $1/L \ll \epsilon \ll m$ to access the desired amplitudes. However, the 
perturbative example of Sec.~\ref{s:pt} indicates that known aspects of the 
$\epsilon$-dependence may lead to a milder extraction problem than in the case 
of total rates.

A final important distinction compared to Ref.~\cite{Hansen:2017mnd} 
is that in contrast to a renormalized current, a single-particle interpolation field is not uniquely defined. This is treated in the present approach exactly as in the LSZ formalism, by  approaching  the pole and dividing out the operator overlap factor. The freedom to choose an  interpolating operator may thus be used to give better constraints on the extraction of the amplitude. To this end it is important to note that in the example above the second pion interpolator should not
be optimized to maximally overlap a single finite-volume state. In contrast, overlapping multiple states is crucial as it is the sum over these that leads to the estimated scattering amplitude. 

Preceding Ref.~\cite{Hansen:2017mnd}, prospects for extracting the hadronic tensor by solving the inverse problem on a Euclidean four-point function are
discussed and tested in   Refs.~\cite{Liu:1993cv,Liu:2016djw,Liu:2017lpe}. In contrast to Ref.~\cite{Hansen:2017mnd} and this article, these earlier 
publications do not explore the role of smearing nor of finite-volume effects. 
Another interesting direction is discussed in Ref.~\cite{Hashimoto:2017wqo}, 
which details an exploratory lattice study of Euclidean four-point functions  relevant for  semi-leptonic $B$-decays. Ref.~\cite{Hashimoto:2017wqo} advocates avoiding the inverse problem by integrating experimental data against a multi-pole function to extract moments that can be directly compared to lattice data.

It is also worth considering the relation of the present work to Ref.~\cite{Agadjanov:2016mao}, which describes a method for extracting the optical potential from a lattice calculation. The approach of Ref.~\cite{Agadjanov:2016mao} requires reconstructing the finite-volume optical potential, known to contain an infinite tower of poles, by fitting a function of this form, applying an $i \epsilon$-prescription, and finally estimating the limit $L \to \infty$ followed by $\epsilon \to 0$. The ordered double limit is a common feature but this earlier work differs in that it is based on measuring finite-volume energies (with twisted boundary conditions) and inserting the $i \epsilon$ via a fit function.

Finally, Ref.~\cite{Barata:1990rn}  considers the extent to which scattering amplitudes can be reconstructed from Euclidean field theories, but from a formal perspective. That work is motivated by conceptual problems that originate in the nonuniqueness of discretization effects together with the role of the continuum limit in defining the analytic continuation to Minkowski signature. These issues are avoided here by expressing the relation through the finite-volume spectral function.

\bigskip

As explained in Sec.~\ref{s:met}, an arbitrary
$m$-to-$n$ particle amplitude requires a  (complex) spectral function with 
$r=m+n-2$ energy arguments,  with one fewer argument 
if no external current is included and one more if either $m$ or $n$ is zero. Proposals for efficiently isolating the scattering amplitude within the spectral function are discussed in Sec.~\ref{s:calc} although no 
numerical results are presented. 
For simplicity of notation, the method is illustrated for a single component 
real scalar field.  

 A number of different approaches can be used to determine 
 $\hat{\rho}_{\boldsymbol{p}_1\boldsymbol{p}_1}^{L,J,\epsilon}(E) $ from 
 $C_{\boldsymbol{p}_1\boldsymbol{p}_1}^{ L, J}(\tau)$ and this work is 
 agnostic to the method employed. It is however crucial that the resolution 
 function, $\hat{\delta}_{\epsilon}(\omega,E)$, approximates the specific complex 
 pole form. 
Sec.~\ref{s:calc} includes a brief discussion of the various possibilities, including the use of a variational method to explicitly build up the spectral function by individually extracting the finite-volume energies and matrix elements. Some probable first applications of the 
general formalism are 
presented in Sec.~\ref{s:appl}. Sec.~\ref{s:pt} details 
a perturbative test elucidating the 
interplay between real and imaginary time, the finite volume, the smearing kernel, and the $i\epsilon$-prescription. The outlook and 
conclusions are in Sec.~\ref{s:concl} together with some brief remarks on the 
straightforward
generalization to multiple species of complex arbitrary spin
fields. 


\section{LSZ reduction}\label{s:met}

This section details the
somewhat non-standard application of the LSZ reduction approach used in this 
work and is therefore restricted to infinite-volume Minkowski correlators. 
The main result given in 
Eq.~\ref{e:big} is the relation between 
a smeared spectral function with a particular smearing kernel and 
the exclusive $m$-to-$n$ scattering amplitude for the process 
$\boldsymbol{p}_1 + \cdots + \boldsymbol{p}_m + J \rightarrow \boldsymbol{p}_{m+1} + \cdots + \boldsymbol{p}_{m+n}$ which is mediated by the external local current
$\hat{J}(x)$.\footnote{If the current has vacuum quantum numbers, it is assumed that the vacuum expectation value is subtracted.} It is assumed that $m$ and $n$ are non-zero for clarity of this general presentation. Zero-to-two transitions are discussed in Sec.~\ref{s:0to2}. The expressions for a purely hadronic process with no external current are provided at the end of the section, culminating in the analogous main result in Eq.~\ref{e:big2}.

For simplicity of notation consider an arbitrary theory with a single 
real scalar field. The LSZ formalism~\cite{LSZ} holds under several 
assumptions including the existence of a mass gap, 
complicating the application to lattice QCD+QED unless 
the photon is given a non-zero mass.   
Based on the asymptotic conditions of Haag-Ruelle scattering theory~\cite{Haag,Ruelle}, 
there is some 
generality in the $n$-point functions to which the LSZ 
reduction is applied. The LSZ procedure is conventionally applied to connected  
time-ordered Feynman functions 
\begin{gather}\label{e:ffun}
	\widetilde{G}_{\rm c}^{J}(q_1,\dots,q_{m+n}) = \int \prod_{j=1}^{m+n} \left\{ d^4x_j \, {\rm e}^{-iq_j\cdot x_j}\right\} \langle 0 | T \left\{ 
	\hat{\phi}(x_1)\cdots   
	\hat{\phi}(x_{m+n}) 
	\hat{J}(0)\right\}|0 \rangle_{\rm c}	 \,,
\end{gather}
where $q_j\cdot x_j = q_j^0t_j - \boldsymbol{q}_j\cdot \boldsymbol{x}_j$, $\hat{\phi}(x)$ is a suitable interpolating operator for a 
single-component real scalar field, and $\boldsymbol{q}_j = \pm\boldsymbol{p}_j$. 
 The integrals defining the Fourier transform are well-defined (once an ultraviolet regularization is put in place) because of the implicit $i \epsilon$-prescription in the Hamiltonian, $\hat H \to \hat H - i \epsilon$.
 This same prescription can be encoded in the momentum coordinates
 via the replacement $q_j^0 \rightarrow  q_j^0 \pm i\epsilon$, which is applied in all temporal Fourier transforms. 
 The fully-connected amplitude 
${\cal M}_{\rm c}^{J}(p_{m+n}\cdots p_{m+1}|p_{m}\cdots p_1)$ (where the $\{p_j\}$ are on-shell four momenta) is  then obtained from 
the residue of the pole in $\widetilde{G}_{\rm c}^J(q_1,\dots,q_{m+n})$ as $q_j^0 \rightarrow \pm E(\boldsymbol{p}_j)$.  The 
upper (lower) signs for both the energy-momentum coordinates and the $i \epsilon$ are taken for incoming (outgoing) particles, respectively. 

Rather than using spectral functions from the time-momentum representation of the $(m+n+1)$-point Feynman functions in Eq.~\ref{e:ffun}, we instead 
define alternative $(m+n-1)$-point functions in which the integration runs over a single time ordering. The latter are referred to as `endcap functions'  
\begin{multline}\label{e:toef}
	\widetilde{F}_{\boldsymbol{p}_{m+n}\boldsymbol{p}_1}^{J} (q_{r+1}, \dots, q_{2}) =  \int \prod_{j=2}^{r+1} \left\{ d^4x_j \, {\rm e}^{-iq_j\cdot x_j}\right\}
	\, \theta(t_{r+1} - t_{r}) \cdots \theta(t_{m+1}) \,  
	\\[-5pt] \times
	\theta(-t_m)\cdots \theta(t_3-t_2) \, 
	\langle \boldsymbol{p}_{m+n} | \hat{\phi}(x_{r+1}) \cdots \hat{\phi}(x_{m+1}) \,
	\hat{J}(0) \hat{\phi}(x_m) \cdots \hat{\phi}(x_2) | \boldsymbol{p}_1 \rangle_{\rm c}   \,,
\end{multline}
where $| \boldsymbol{p} \rangle$ denotes a one-particle state with the usual normalization $\langle \boldsymbol p' \vert \boldsymbol p \rangle = 2 E(\boldsymbol p) (2 \pi)^3 \delta^3(\boldsymbol p' - \boldsymbol p)$, and 
$r = (m+n-2)$  Heaviside step functions are used to enforce a particular time ordering.
 Note that it would make no difference to include a time ordering operator in the matrix element in this expression.
Spectral functions introduced from the time-momentum representations of endcap 
functions have $r$ arguments, rather 
than the $r+2$ required for Feynman functions, and thus are presumably more 
amenable to numerical determination. Furthermore, the selection of a single 
time ordering requires only a single spectral function  whereas one is required 
for each of the time orderings in Eq.~\ref{e:ffun}. 
The full Lorentz covariance of the Feynman functions is lost when the endcaps 
and  
a single time ordering are employed, but is of course recovered in the scattering amplitude.  

Although the analytic structure of the endcap functions in Eq.~\ref{e:toef}
is different than that of the Feynman functions of Eq.~\ref{e:ffun}, 
it is demonstrated in App.~\ref{a:lsz} that LSZ reduction of $r$ interpolating 
fields yields the same on-shell pole  
\begin{multline}\label{e:tlsz}
	\widetilde{F}_{\boldsymbol{p}_{m+n}\boldsymbol{p}_1}^{J} (q_{m+n-1}, \dots, q_{2}) = 
	\prod_{j = 2}^{r+1}\left\{ \frac{Z^{1/2}(\boldsymbol{p}_j) }{2E(\boldsymbol{p}_j)}\right\} \, {\cal M}_{\rm c}^{J}(p_{m+n}\cdots p_{m+1}|p_{m} \cdots p_1) \,  
	\\ 
	\times \prod_{j=2}^{m} \left\{ \frac{i}{q_j^0 - E(\boldsymbol{p}_j) + i\epsilon}\right\}  
	\prod_{k=m+1}^{r+1} \left\{ \frac{i}{-q_k^0 - E(\boldsymbol{p}_k) + i\epsilon}\right\} \, + \cdots   \,,
\end{multline}
where $Z^{1/2}(\boldsymbol{p}) = \langle \boldsymbol{p} | \hat{\phi}(0) | 0 \rangle$ and `$+\cdots$' denotes additional contributions which are regular near 
the pole. 
The endcap function in Eqs.~\ref{e:toef} and~\ref{e:tlsz} can also be expressed as a spectral function 
\begin{equation}\label{e:tspec} 
\widetilde{F}_{\boldsymbol{p}_{m+n}\boldsymbol{p}_1}^{J} (q_{r+1}, \dots, q_{2}) =  \hat{\rho}_{\boldsymbol{p}_{m+n}\boldsymbol{p}_1}^{J,\epsilon}(q_{r+1}, \dots, q_2)  \equiv \int_0^\infty \! \! \frac{d^r\! E}{\pi^r}  \  \hat{\delta}^r_{\epsilon}(q^0,E)  \, \rho_{\boldsymbol{p}_{m+n}\boldsymbol{p}_1}^{J} \! \big (
	(E_{r+1},-\boldsymbol{p}_{r+1}), \dots,(E_2,\boldsymbol{p}_2 ) \big) \,,
\end{equation}
where
\begin{align}  
\int_0^\infty \! \! \frac{d^r\! E}{ \pi^r} &  \equiv \prod_{j=2}^{r+1} \left\{ \int_0^{\infty} \frac{dE_ j}{\pi} \right\} \,, \\
\label{e:tspecDELTA}
\hat{\delta}^r_{\epsilon}(q^0,E) &  \equiv   \prod_{k = m+1}^{r+1} \left\{ \frac{i}{E(\boldsymbol{p}_{m+n}) - \sum_{l=k}^{r+1} q_l^0 - E_k + i\epsilon}\right\} 	\times \prod_{k = 2}^{m} \left\{ \frac{i}{\sum_{l=2}^{k} q_l^0 + E(\boldsymbol{p}_1) - E_k + i\epsilon}\right\} \,, \\
\begin{split}
	\rho_{\boldsymbol{p}_{m+n}\boldsymbol{p}_1}^{J}(k_{r+1}, \dots, k_{2}) & \equiv 
	\sum_{\alpha_{r+1},\dots,\alpha_2} \prod_{j=2}^{r+1}\left\{ \pi \delta(E_{\alpha_j} -
	k_j^0)\right\} \times 
	\langle \boldsymbol{p}_{m+n} | \hat{\varphi}(\boldsymbol{k}_{r+1},0) | \alpha_{r+1} \rangle \times \cdots  \times \\[-5pt]
	& \hspace{150pt} \times	
	\langle \alpha_{m+1} | \hat{J}(0) | \alpha_m \rangle \times  \cdots  \times
	\langle \alpha_2 | \hat{\varphi}(\boldsymbol{k}_2,0) | \boldsymbol{p}_1 \rangle  \, - \cdots   \,. 
	\end{split}   
\end{align}
 Here $\hat{\varphi}(\boldsymbol{p},t) = \int d^3\boldsymbol{x} \, {\rm e}^{i\boldsymbol{p}\cdot \boldsymbol{x}} \hat{\phi}(\boldsymbol{x},t)$, $E_{\alpha}$ is the energy of the state $|\alpha\rangle$, and the ellipsis denotes disconnected contributions which must be subtracted explicitly. 
The sums of states indexed by $\alpha_j$ are shorthand for individual integrals 
over each of the fixed-particle-number sectors of the Hilbert space of 
states, formally defined as~\cite{Hansen:2017mnd}
\begin{gather}
	\sum_{\alpha} \equiv \sum_{n_{\alpha}} \frac{1}{n_{\alpha}!} \int \prod_{j=1}^{n_{\alpha}} d\Gamma(\boldsymbol{k}_j)  \,,
\end{gather}
where $d\Gamma(\boldsymbol{k}) \equiv \frac{d^3\boldsymbol{k}}{(2\pi)^3 \, 
2E(\boldsymbol{k})}$ is the usual Lorentz-invariant integration measure.
Because these relations are employed at finite $\epsilon$, the particular form of 
the pole factors introduced above must be maintained throughout and not manipulated using infinitesimal-$\epsilon$ identities.  

\bigskip

 As already suggested by the notation, the central idea of our proposal is to view the pole factors in Eq.~\ref{e:tspecDELTA} as complex smearing 
functions akin to  $\hat{\delta}_{\epsilon}(E,\omega)$ in Eq.~\ref{e:sp2}. 
Eq.~\ref{e:tspec} thus defines the smeared spectral function $\hat{\rho}_{\boldsymbol{p}_{m+n}\boldsymbol{p}_1}^{J,\epsilon}(q_{r+1}, \dots, q_2)$ with characteristic width $\epsilon$. Based on Eq.~\ref{e:tlsz}, the desired amplitude is obtained from the limit 
\begin{gather}\label{e:big} 
{\cal M}_{\rm c}^{J}(p_{m+n}\cdots p_{m+1}| p_{m}\cdots p_1) = 
	\prod_{j=2}^{r+1} \frac{2E(\boldsymbol{p}_j)}{Z^{1/2}(\boldsymbol{p}_j)} \, 
	\times 
	\lim_{\epsilon \rightarrow 0^{+}}\, \epsilon^{r}  \hat{\rho}_{\boldsymbol{p}_{m+n}\boldsymbol{p}_1}^{J,\epsilon} \big ((-E(\boldsymbol{p}_{r+1}), -\boldsymbol{p}_{r+1}), \dots, 
	(E(\boldsymbol{p}_2),\boldsymbol{p}_2) \big)  \,,
\end{gather}
 where the arguments of the smeared spectral function have been set on shell 
and the poles in Eq.~\ref{e:tlsz} amputated by the factor $\epsilon^r$.
Practically, the smeared spectral function at finite $\epsilon$,  
$\hat{\rho}_{\boldsymbol{p}_{m+n}\boldsymbol{p}_1}^{J,\epsilon}(q_{r+1}, \dots, q_2)$, is obtained from the $L\rightarrow \infty$ limit of finite-volume 
spectral functions with a particular smearing kernel. 
The relation in Eq.~\ref{e:big} between amplitudes and smeared spectral functions in the $\epsilon \rightarrow 0^{+}$ limit 
is a central tenet of the method. 
The overlap factors $Z(\boldsymbol{p})$ are in fact independent of $\boldsymbol{p}$ due to Lorentz covariance but their 
momentum dependence is retained as a notational reminder that arbitrary 
interpolators may be used for each of the particles.
These overlap 
factors do acquire momentum dependence if calculated in a
finite volume or if 
a spatial smearing wavefunction is introduced. 

 For completeness, we close by giving the formulae for
endcap functions $\widetilde{F}_{\boldsymbol{p}_{m+n}\boldsymbol{p}_1}(q_{r+1},\dots,q_2)$
that do not contain an external current. The LSZ reduction in Eq.~\ref{e:tlsz}
now contains an overall four-momentum conserving $\delta$-function which we 
omit 
from the definition of the spectral function 
\begin{gather}\label{e:amp2}
	\widetilde{F}_{\boldsymbol{p}_{m+n}\boldsymbol{p}_1}(q_{r+1},\dots,q_2) = 
	(2\pi)^4 \delta^4(p_{m+n} - q_{r+1} - \cdots - q_2 - p_1) \, \hat{\rho}^{\epsilon}_{\boldsymbol{p}_{m+n}\boldsymbol{p}_1}(q_{r+1},\dots,q_{m+2},q_{m}, \dots,
	q_2)\,, 
\end{gather}
where the smeared spectral function has now has $r-1$ energy-momentum 
arguments chosen to omit $q_{m+1}$. The smeared and unsmeared
spectral functions are given by 
\begin{align}
\begin{split}
\hat{\rho}^{\epsilon}_{\boldsymbol{p}_{m+n}\boldsymbol{p}_1}(q_{r+1},\dots,q_{m+2},q_{m}, \dots,
	q_2) &= \int_0^{\infty}\,  \frac{d^{r-1}E}{\pi^{r-1}} 
	\hat{\delta}^{r-1}_{\epsilon}(q^0,E)  
	\\ 
	&   \times \rho_{\boldsymbol{p}_{m+n}	\boldsymbol{p}_1}\big(
	(E_{r+1},-\boldsymbol{p}_{r+1}), \dots,(E_{m+2},\boldsymbol{p}_{m+2}),(E_{m},\boldsymbol{p}_{m}),\dots,
	(E_{2},\boldsymbol{p}_{2}) \big) \,,
	\end{split}
	\\[5pt]
\begin{split}
	\rho_{\boldsymbol{p}_{m+n}\boldsymbol{p}_1}(k_{r+1}, \dots,k_{m+2},k_{m+1},\dots, k_{2}) & \equiv 
	\sum_{\alpha_{r+1},\dots,\alpha_{m+2},\atop \alpha_m,\dots,\alpha_2} \prod_{j=2, \atop j\ne m+1}^{r+1}\left\{ \pi \delta(E_{\alpha_j} -
	k_j^0)\right\} \times 
	\langle \boldsymbol{p}_{m+n} | \hat{\varphi}(\boldsymbol{k}_{r+1},0) | \alpha_{r+1} \rangle  \\[0pt]
	& \hspace{70pt} \times \cdots \times	
	\langle \alpha_{m+2} | \hat{\phi}(0) | \alpha_{m} \rangle \times  \cdots  \times
	\langle \alpha_2 | \hat{\varphi}(\boldsymbol{k}_2,0) | \boldsymbol{p}_1 \rangle \, - \cdots    \,, 
	\end{split}
\end{align}
where the disconnected contributions are again subtracted and the integration measure $\frac{d^{r-1}E}{\pi^{r-1}}$ omits $E_{m+1}$. Similarly, the smearing kernel $\hat{\delta}^{r-1}_{\epsilon}(q^0,E)$ is 
defined as in Eq.~\ref{e:tspecDELTA} but with $\prod_{k=m+2}^{r+1}$ in the first factor. 

The amplitude is recovered as in Eq.~\ref{e:big}
\begin{multline}\label{e:big2} 
i{\cal M}_{\rm c}(p_{m+n}\cdots p_{m+1}| p_{m}\cdots p_1) = 
	\prod_{j=2}^{r+1} \frac{2E(\boldsymbol{p}_j)}{Z^{1/2}(\boldsymbol{p}_j)} \, 
	\\ 
	\times
	\lim_{\epsilon \rightarrow 0^{+}}\, \epsilon^{r}  \hat{\rho}_{\boldsymbol{p}_{m+n}\boldsymbol{p}_1}^{\epsilon}\big ((-E(\boldsymbol{p}_{r+1}), -\boldsymbol{p}_{r+1}), \dots, 
	(-E(\boldsymbol{p}_{m+2}),-\boldsymbol{p}_{m+2}),(E(\boldsymbol{p}_{m}),\boldsymbol{p}_{m}),\dots,
	(E(\boldsymbol{p}_{2}),\boldsymbol{p}_{2}) \big)  \,,
\end{multline}
with the conventional factor of $i$ included in its definition. Note that 
although $r-1$ pole factors have been introduced in the smearing kernel, the LSZ reduction necessitates that $r$ such poles must be amputated at the on-shell point. This  subtlety  is discussed further in Sec.~\ref{s:pt}.

\section{Calculation of the spectral functions}\label{s:calc}

The smeared spectral function defined in Eq.~\ref{e:tspec} can be obtained 
from finite-volume Euclidean lattice simulations. To this end, we employ 
connected $(m+n+1)$-point\footnote{We again assume $m$ and $n$ are non-zero for clarity.} and two-point Euclidean correlation functions in the time-momentum representation 
\begin{align}\label{e:ecor}
	C^{J,L}_{(m+n+1){\rm pt}} \big ((-\boldsymbol{p}_{n+m},\tau_{n+m}), \dots, (\boldsymbol{p}_1,\tau_1) \big) & = 
	{\langle} 0 | \hat{\varphi}(-\boldsymbol{p}_{n+m},\tau_{n+m}) \dots 
	\hat{J}(0) \dots \hat{\varphi}(\boldsymbol{p}_1,\tau_1) | 0 \rangle_{{\rm c},L} \,,
\\
	C^{L}_{2{\rm pt}}(\boldsymbol{p},\tau) & = { \langle} 0 | \hat{\varphi}(\boldsymbol{p},\tau) 
	\hat \phi(0) | 0 \rangle_{{\rm c},L} \,,
\end{align}
where $\tau_{n+m} > \tau_{n+m-1} > \cdots > \tau_{m+1} > 0 > \tau_{n} > \cdots > \tau_1$, and    
$\hat{\varphi}(\boldsymbol{p},\tau) = \int_{\Omega_L} d^3\boldsymbol{x} \,{\rm e}^{i
\boldsymbol{p}\cdot \boldsymbol{x}} \hat{\phi}(x)$ where $\Omega_L$ denotes the periodic three-torus of size $L$. Note that only finite-volume momenta 
$\boldsymbol{p} = 2\pi\boldsymbol{n}/L$, where $\boldsymbol{n}$ is a vector of integers, are possible. 
These correlation 
functions are amenable to calculation with lattice simulations, and are used 
to define Euclidean endcap functions
\begin{align}\label{e:eef}
\begin{split}
	C^{J,L}_{\boldsymbol{p}_{n+m}\boldsymbol{p}_1} \big ((-\boldsymbol{p}_{r+1}, \tau_{r+1}) , \dots, (\boldsymbol{p}_2,\tau_2) \big) & = 2 \sqrt{E (\boldsymbol{p}_{n+m}) E (\boldsymbol{p}_1)} L^3
	Z^{1/2}(\boldsymbol{p}_{n+m}) 
	Z^{1/2}(\boldsymbol{p}_1) \, \\
& \hspace{50pt}  \times	\lim_{{\tau_{n+m} - \tau_{r+1},
	\atop {
		\tau_2-\tau_1 }} \rightarrow \infty}
	\frac{C^{J,L}_{(m+n+1){\rm pt}}\big ((-\boldsymbol{p}_{n+m},\tau_{n+m}), \dots, (\boldsymbol{p}_1,\tau_1) \big )}{C^{L}_{2{\rm pt}}(\boldsymbol{p}_{m+n},\tau_{m+n} - \tau_{r+1})\,
		C^{L}_{2{\rm pt}}(\boldsymbol{p}_{1},\tau_{2} - \tau_{1}   )} \,,
		\end{split}
		\\[8pt] &  \hspace{-100pt} =    2 \sqrt{E (\boldsymbol{p}_{n+m}) E (\boldsymbol{p}_1)} L^3
		{ \langle} \boldsymbol{p}_{m+n} | \hat{\varphi}(-\boldsymbol{p}_{r+1},0)\, 
		{\rm e}^{-\hat{H}(\tau_{r+1}-\tau_r)} \cdots \hat{J}(0)\cdots {\rm e}^{-\hat{H}
		(\tau_3-\tau_2)}\hat{\varphi}(\boldsymbol{p}_2,0) | \boldsymbol{p}_1 \rangle_{{
			\rm c},L} \,,
		\\[5pt]\label{e:eef3}
		& \hspace{-100pt} = \int_0^\infty \! \! \frac{d^r\! E}{\pi^r}
		{\rm e}^{-E_{r+1}(\tau_{r+1}-\tau_r)}\cdots {\rm e}^{-E_2(\tau_3-\tau_2)}\, 
	\rho^{L,J}_{\boldsymbol{p}_{m+n}\boldsymbol{p}_1} \big ((E_{r+1},-\boldsymbol{p}_{r+1}),\dots, (E_2,\boldsymbol{p}_2) \big) \,,
\end{align}
where $\hat{H}$ is the finite-volume Hamiltonian and the two outermost fields have been placed on-shell by taking the 
asymptotic time separation limit. Throughout this work all finite-volume states are normalized to unity. A key observation is that the endcap function in Eq.~\ref{e:toef} has a spectral function identical to the one in Eq.~\ref{e:eef3}  in a suitably-defined $L\rightarrow \infty$
limit.
Analytic continuation between the Minkowksi and Euclidean endcap functions 
thus proceeds via the 
spectral function, which is agnostic toward the metric signature.

The $L \to \infty$ limit is made well-defined by convolution with a resolution function, the specific choice of which determines the extracted quantity.   In particular, using the $\hat \delta_\epsilon^r$ defined in the previous section we find 
\begin{equation}
\label{e:slim2} 
	\hat{\rho}^{J,\epsilon}_{\boldsymbol{p}_{m+n},\boldsymbol{p}_1}(q_{r+1}, 
	\dots,q_2) = \lim_{L\rightarrow \infty}
	\hat{\rho}^{L,J,\epsilon}_{\boldsymbol{p}_{m+n},\boldsymbol{p}_1}(q_{r+1}, 
	\dots,q_2),
\end{equation}
where
\begin{align}\label{e:tspecFV} 
   \hat{\rho}_{\boldsymbol{p}_{m+n}\boldsymbol{p}_1}^{L,J,\epsilon}(q_{r+1}, \dots, q_2) &  \equiv \int_0^\infty \! \! \frac{d^r\! E}{ \pi^r}  \  \hat{\delta}^r_{\epsilon}(q^0,E)  \, \rho_{\boldsymbol{p}_{m+n}\boldsymbol{p}_1}^{L,J} \! \big (
	(E_{r+1},-\boldsymbol{p}_{r+1}), \dots,(E_2,\boldsymbol{p}_2 ) \big) \,,
\\
	\begin{split}\label{e:fvun}
		\rho_{\boldsymbol{p}_{m+n}\boldsymbol{p}_1}^{L,J}(k_{r+1}, \dots, k_{2}) & \equiv 2 \sqrt{E(\boldsymbol p_{m+n}) E(\boldsymbol p_{1})} L^3
	\sum_{n_{r+1},\dots,n_2} \prod_{j=2}^{r+1}\left\{ \pi \delta(E^L_{n_j} -
	k_j^0)\right\} \times 
	\langle \boldsymbol{p}_{m+n} | \hat{\varphi}(\boldsymbol{k}_{r+1},0) | n_{r+1} \rangle_L \times \\ & \hspace{100pt} \times \cdots    %
	\langle n_{m+1} | \hat{J}(0) | n_m \rangle_L \times  \cdots  \times
	\langle n_2 | \hat{\varphi}(\boldsymbol{k}_2,0) | \boldsymbol{p}_1 \rangle_L    \,-\cdots\,  
	\end{split}   
\end{align}
and the disconnected contributions to $\rho^{L,J}_{\boldsymbol{p}_{m+n}\boldsymbol{p}_1}$ are subtracted explicitly.  
 As $L$ increases, 
	the allowed finite-volume three-momenta will change. However, this difference 
	becomes irrelevant as $L\rightarrow \infty$ due to
	the increasing density of states. The infinite-volume 
	limit in Eq.~\ref{e:slim2} formally assumes that 
	nearly equivalent finite-volume momenta are selected at each $L$.

\bigskip

		So far we have not discussed algorithms for solving the inverse problem and 
	determining the smeared spectral function $\hat{\rho}^{L,J,\epsilon}_{\boldsymbol{p}_{m+n},
	\boldsymbol{p}_1}$ from the Euclidean endcap functions $C^{J,L}_{\boldsymbol{p}_{n+m}\boldsymbol{p}_1}$. Much work in this direction has already been performed in  both zero- and non-zero temperature lattice QCD, but typically with the 
	aim of achieving a sharply peaked resolution function. Applications 
	of the approach described here will require extensive numerical tests which
	are deferred to future work. 
Here we only comment briefly on some possible strategies.

First, the Backus-Gilbert method, discussed in Refs.~\cite{BG1,BG2}, gives a linear, 
model-independent estimator of a smeared spectral function in which the 
covariance matrix of the data is used to stabilize the inverse. 
The algorithm was examined in the context of total hadronic rates in 
Ref.~\cite{Hansen:2017mnd} where the aim is to achieve a narrow 
resolution function with unit area, as indicated in Eq.~\ref{e:sp2}. 
Ref.~\cite{Hansen:2017mnd} stresses that a perfect inverse, in which the 
resolution function becomes arbitrarily close to a $\delta$-function, 
is undesirable 
as it would lead to an extraction dominated by finite-$L$ effects. Instead, 
for a given box size, an optimal width exists that is larger than 
$1/L$ but, in the ideal situation, smaller than the scales over which the 
spectral function varies. 
	In the present case, the desired resolution function 
	is instead a complex pole form
	that is ultimately
	incorporated into the LSZ procedure.

	In fact, as has been recently stressed in 
	Ref.~\cite{Hansen:2019idp}, the Backus-Gilbert algorithm can be modified such 
	that a target resolution function is viewed as an additional input. In this approach, the inverse is chosen to minimize the distance to the target function, rather than to minimize the width. This perspective is highly compatible with the method advocated here, allowing one to directly target the function $\hat{\rho}^{L,J,\epsilon}_{\boldsymbol{p}_{m+n},\boldsymbol{p}_1}$ with a specified value of $\epsilon$. Of course, any linear inverse method must face inherent limitations for a given input data quality. With this in mind we highlight two key features that might make the inverse problems considered here somewhat easier to treat.
	\begin{itemize}
		\item In contrast to the total rates in 
			Ref.~\cite{Hansen:2017mnd}, the present method allows one to use a  large   basis of operators in the definition of the spectral function. These will all differ at finite $\epsilon$ but must coincide in the ordered double limit. Performing a  constrained extrapolation   to a    set of spectral functions may better 
			 determine  the target observable.
		\item The complex pole smearing function introduces an $\epsilon$-dependence with useful analyticity properties. In particular, amplitudes must have a convergent expansion in this parameter, with radius of convergence limited by the nearest branch point. This is in contrast to gaussian smearing kernels $ \hat{\delta}_{\epsilon}(E,\omega) \sim \exp{[- (E - \omega)^2/\epsilon^2]}$, which exhibit an essential singularity at $\epsilon = 0$. To this end note that for total-rate applications, the imaginary part of the complex pole ($\sim \epsilon/[(E - \omega)^2 + \epsilon^2]$) may provide a more useful finite-width 
			$\delta$-function approximation.
	\end{itemize}
These features should allow one to maximally constrain the scattering amplitude without demanding too much resolution in the inverse.

In addition to Backus-Gilbert there are 
	a number of other approaches to this well-known `unfolding' problem due to its ubiquity. 
	For example, the Maximum-Entropy Method (MEM) defines a likelihood function to determine the most probable solution for the available data set. To do so one requires a starting ansatz, referred to as the default model or prior estimate.  In any finite-volume application, MEM clearly performs some smearing as the true finite-volume spectral function is a series of Dirac $\delta$-functions. However, in contrast to the Backus-Gilbert approach, the form of the resolution function is less clear.\footnote{Both the Backus-Gilbert algorithm and the Maximum-Entropy Method have also been applied to finite-temperature lattice QCD~\cite{Asakawa:2000tr,Brandt:2015sxa}.} 
	
	Finally we emphasize that all finite-volume spectral functions are defined 
	from finite-volume energies and matrix elements. These can, 
	in principle, be accessed individually by employing solutions of 
	Generalized Eigenvalue Problems (GEVP) using a large basis of interpolating 
	operators~\cite{Michael:1985ne,Luscher:1990ck,Blossier:2009kd}. In this 
	approach it is important to distinguish between the 
	interpolating fields used to create the individual scattering particles, and 
	the optimized operators used to access the finite-volume states. The latter 
	do not enter the spectral function as they are divided out in determining 
	finite-volume matrix elements, while the former appear as `current insertions' which are put on the single-particle mass shell by selecting the appropriate four-momentum transfer.  This exact finite-volume reconstruction is performed by
	using Eq.~\ref{e:fvun} in Eq.~\ref{e:tspecFV} to obtain
\begin{multline}\label{e:fvol}
	\hat{\rho}_{\boldsymbol{p}_{m+n}\boldsymbol{p}_1}^{L,J,\epsilon}(q_{r+1}, \dots, q_2) = 2 \sqrt{E(\boldsymbol{p}_{m+n})E(\boldsymbol{p}_1)}L^3  \sum_{n_{r+1}, 
		\dots, n_2 } \hat{\delta}_{\epsilon}^{r}(q^0, E^L_n)\, \langle \boldsymbol{p}_{m+n} | \hat{\varphi}(\boldsymbol{k}_{r+1},0) | n_{r+1} \rangle_L \,\times \\
		 \times \cdots  \times 
		\langle n_{m+1} | \hat{J}(0) | n_m \rangle_L \times  \cdots  \times
	\langle n_2 | \hat{\varphi}(\boldsymbol{k}_2,0) | \boldsymbol{p}_1 \rangle_L    \,- \cdots\,,  
\end{multline}
	where the disconnected contributions are subtracted and $\hat{\delta}_{\epsilon}^{r}(q^0, E^L_n)$ is short-hand for 
	the smearing kernel of Eq.~\ref{e:tspecDELTA} with arguments 
$\{q^0_{r+1},\dots,q_2^0\}$ and $\{E^L_{n_{r+1}},\dots,E^L_{n_2}\}$. 
Practical applications of this finite volume 
reconstruction treat a few terms in the sum over 
$\{ n_{r+1},\dots,n_2\}$ and are considered in Sec.~\ref{s:appl}. Such 
applications require  
the finite-volume energies and matrix elements shown in 
Eq.~\ref{e:fvun}. Since the smearing kernel is peaked for small $\epsilon$, it is likely that only finite-volume states with energies 
$\{E^L_{n_j}\}$ near the on-shell point are important. 
	
	\bigskip
	
In addition to various reconstruction methods one can take advantage of known properties 
of the spectral function to potentially ameliorate the solution of this inverse problem. 
For example, it is likely that solving real inverse problems is preferable to the 
complex one in Eq.~\ref{e:slim2}. To this end, the real and imaginary parts of 
$\hat{\rho}^{J,\epsilon}_{\boldsymbol{p}_{m+n},\boldsymbol{p}_1}(q_{r+1}, 
	\dots,q_2)$ can be determined separately from solutions of 
	four inverse problems.
	 To see this note that the real and imaginary parts of 
	 $\hat{\rho}_{\boldsymbol{p}_{m+n}\boldsymbol{p}_1}^{L,J,\epsilon}$ can be 
	 trivially divided as follows
	\begin{align}\label{e:reim}
	\text{Re} \, \hat{\rho}_{\boldsymbol{p}_{m+n}\boldsymbol{p}_1}^{L,J,\epsilon}(q_{r+1}, \dots, q_2) &  = \int_0^\infty \! \! \frac{d^r\! E}{ \pi^r}  \Big [ 	\text{Re} \,  \hat{\delta}^r_{\epsilon}(q^0,E)  \, 	\text{Re} \, \rho_{\boldsymbol{p}_{m+n}\boldsymbol{p}_1}^{L,J} \!   ( \, \cdots ) - \text{Im} \,  \hat{\delta}^r_{\epsilon}(q^0,E)  \, 	\text{Im} \, \rho_{\boldsymbol{p}_{m+n}\boldsymbol{p}_1}^{L,J} \!   ( \, \cdots ) \Big ] \,,
	\\
	\text{Im} \, \hat{\rho}_{\boldsymbol{p}_{m+n}\boldsymbol{p}_1}^{L,J,\epsilon}(q_{r+1}, \dots, q_2) &  \equiv \int_0^\infty \! \! \frac{d^r\! E}{ \pi^r}  \Big [ 	\text{Re}  \, \hat{\delta}^r_{\epsilon}(q^0,E)  \, 	\text{Im} \, \rho_{\boldsymbol{p}_{m+n}\boldsymbol{p}_1}^{L,J} \!   (\, \cdots ) + \text{Im} \,  \hat{\delta}^r_{\epsilon}(q^0,E)  \, 	\text{Re} \, \rho_{\boldsymbol{p}_{m+n}\boldsymbol{p}_1}^{L,J} \!   ( \, \cdots ) \Big ] \,,
	\end{align}
 where we have suppressed the arguments of the unsmeared spectral functions on the right-hand side. 
 The four convolution integrals on the right-hand side  define the solutions to four inverse problems, each defined using only the real or imaginary part of the Euclidean endcap function. This follows since the real and imaginary parts of this endcap function contain 
only the real and imaginary parts of the unsmeared spectral function, respectively. 

	Finally, we point out a property of the spectral functions which may 
	further aid their numerical determination. Employing a `unitarity cut' and 
	inserting a complete set of (finite-volume) states gives 
	\begin{gather}\label{e:ucut}
		\hat{\varrho}^{L,J,\epsilon}_{\boldsymbol{p}_{m+n},\boldsymbol{p}_1}(q_{r+1}, 
		\dots,q_2) = \sum_{n_j} \delta_{\boldsymbol{p}_{\alpha_j},\boldsymbol{p}_{j-1} + \dots + \boldsymbol{p}_1} 
		\frac{
		i\, \hat{\varrho}^{L,J,\epsilon}_{\boldsymbol{p}_{m+n},n_j} ( q_{r+1}, \dots,q_{j+1})
		\,  
		\hat{\varrho}^{L,\epsilon}_{n_j,\boldsymbol{p}_1} ( q_{j-1}, \dots,q_{2}) }{\sum_{l=2}^j q_l^0 + E(\boldsymbol{p}_1) - E^L_{n_j} + i\epsilon} \,,
	\end{gather}
 where $j<m$ and  $\hat{\varrho}^{L,J,\epsilon}_{\boldsymbol{p}_{m+n},n_j}$ denotes the spectral function of the full correlator (with $|n_j\rangle_L$ as the right endcap and the appropriate normalization)  including disconnected contributions. 
	Based on this relation, the finite-volume smeared spectral function at low energies (where few states contribute) may 
	be partially or fully reconstructed from finite-volume 
	energies and matrix elements. The unitarity cut and finite-volume reconstruction approaches are similar 
	in spirit to Ref.~\cite{DellaMorte:2017khn} which uses the low-lying 
	spectrum and matrix 
	elements from Ref.~\cite{Andersen:2018mau} to reconstruct the vector-vector 
	correlator.

\section{Example applications}\label{s:appl}

A selection of specific applications of the general formalism discussed in 
Sec.~\ref{s:met} are presented here. 
These applications all rely on a multi-particle LSZ reduction, since at least two particles appear in either the initial or final state.  This is in contrast to Ref.~\cite{Hansen:2017mnd}, which describes zero-to-$X$ and one-to-$X$ inclusive rates that do not require this approach.  

\subsection{Zero-to-two transitions}\label{s:0to2}

As a minimal illustration of the LSZ reduction technique, we first consider the zero-to-two exclusive process 
$J \rightarrow \boldsymbol{p}_1 + \boldsymbol{p}_2$ mediated by an external 
current $\hat{J}$. An example of such a process is the timelike pion form 
factor in which an external electromagnetic current produces two final-state 
	pions. In the elastic region, this quantity may be accessed directly from a 
	two-point function by converting a finite-volume two-pion-like state into an actual asymptotic state~\cite{Meyer:2011um}.  The key advantage of the 
	present method is that it holds above arbitrary inelastic thresholds. 
Since lattice calculations of 
two- and three-point temporal correlation functions are well 
established, exclusive zero-to-two transition form 
factors (such as the timelike pion form factor) 
are an ideal first application.

For this process one field is reduced into 
	the asymptotic `out' state using the LSZ procedure. The relevant infinite-volume real-time endcap function, to which the LSZ reduction of App.~\ref{a:lsz} is applied, is given by   
	\begin{align}\label{e:02toef1}
	\widetilde{F}^J_{\boldsymbol{p}_2,0}(q_1) & =  \int d^4x_1 {\rm e}^{-i   q_1\cdot x_1} 
	\theta(t_1) \langle \boldsymbol{p}_2 | \hat{\phi} (x_1) \hat{J}(0) |0\rangle_{\rm c} \,,
\\  
	&= \frac{Z^{1/2}(\boldsymbol{p}_1)}{2E(\boldsymbol{p}_1)} \,	{\cal M}_{\rm c}^J(p_2p_1|0) \frac{i}{-q_1^0 - E(\boldsymbol{p}_1) + 
	i\epsilon} \, + \cdots  \,,
	\\\label{e:02toef3} 
	&= \int_0^{\infty} \frac{dE_1}{\pi} \frac{i}{E(\boldsymbol{p}_2) - q_1^0 - E_1 + i\epsilon}
	\,
	\rho^{J}_{\boldsymbol{p}_2,0}(E_1,-\boldsymbol{p}_1) \equiv \hat{\rho}^{J,\epsilon}_{\boldsymbol{p}_2,0}(q_1)  \,,
\end{align}
	where in Eq.~\ref{e:02toef3} we have expressed the endcap function as a smeared 
spectral function $\hat{\rho}^{J,\epsilon}_{\boldsymbol{p}_2,0}(q_1)$. 
	The matrix element in Eq.~\ref{e:02toef1} has no disconnected contributions, so no such subtractions are necessary.	
	The desired amplitude is obtained as in Eq.~\ref{e:big} 
\begin{gather}
	{\cal M}_{\rm c}^J(p_2p_1|0) = 
\frac{2E(\boldsymbol{p}_1)}{Z^{1/2}(\boldsymbol{p}_1)} 
	\lim_{\epsilon \rightarrow 0^+} \epsilon \, \hat{\rho}_{\boldsymbol{p}_2,0}^{J,\epsilon}(-E(\boldsymbol{p}_1),-\boldsymbol{p}_1) \,,
\end{gather}
by putting the argument of the smeared spectral function on shell. 
While ${\cal M}^{J}_{\rm c}(p_2p_1|0)$ is complex (with a 
phase given by Watson's theorem in the elastic region), four 
real smeared spectral functions may be determined separately 
as described in Eq.~\ref{e:reim}. 
In this procedure, the real and imaginary parts of the 
smearing  
kernel at the on-shell point with $q_1^0=-E(\boldsymbol{p}_1)$ are given by 
\begin{align}\label{e:2ker}
	\hat{\delta}_{\epsilon}(-E(\boldsymbol{p}_1),E_1)	&\equiv 
	\frac{i}{E(\boldsymbol{p}_2) + E(\boldsymbol{p}_1) -E_1 + i\epsilon}  
	=
	\frac{\epsilon}{(E(\boldsymbol{p}_2) + E(\boldsymbol{p}_1) -E_1)^2 + \epsilon^2}
	+ i\frac{E(\boldsymbol{p}_2) + E(\boldsymbol{p}_1) -E_1}{(E(\boldsymbol{p}_2) + E(\boldsymbol{p}_1) -E_1)^2 + \epsilon^2}. 
\end{align}
In order to obtain these spectral functions from finite-volume lattice 
simulations, the required Euclidean endcap function is 
\begin{align}\label{e:02eef}
	C_{\boldsymbol{p}_2,0}^{J,L}(
	-\boldsymbol{p}_1,\tau_1) & = \sqrt{2 E(\boldsymbol p_2) L^3}
	\langle \boldsymbol{p}_2 | \hat{\varphi}(-\boldsymbol{p}_1,0) 
	{\rm e}^{-\hat{H}\tau_1} \hat{J}(0) | 0 \rangle_{{\rm c},L} \,,
\\  
& = 
	Z^{1/2}(\boldsymbol{p}_2) \lim_{\tau_2-\tau_1 \rightarrow \infty}
	\frac{C^{J,L}_{3{\rm pt}} \big ((-\boldsymbol{p}_2,\tau_2),
	(-\boldsymbol{p}_1,\tau_1) \big)}{C^L_{2{\rm pt}}(\boldsymbol{p}_2,\tau_2-\tau_1)}
	= \int_0^{\infty} \frac{dE_1}{\pi} {\rm e}^{-E_1 \tau_1} \rho_{\boldsymbol{p}_2,0}^{L,J}(
	E_1,
	-\boldsymbol{p}_1) \,,
\end{align}
where $\tau_1 > 0$ and the three-point functions
\begin{gather}
	C^{J,L}_{3{\rm pt}}\big ((-\boldsymbol{p}_2,\tau_2),
	(-\boldsymbol{p}_1,\tau_1) \big ) = \langle 0 | 
	\hat{\varphi}(-\boldsymbol{p}_2,0)\,
	{\rm e}^{-\hat{H}(\tau_2-\tau_1)}
	\hat{\varphi}(-\boldsymbol{p}_1,0)\, 
	{\rm e}^{-\hat{H}\tau_1}
	\hat{J}(0) | 0 \rangle_{{\rm c},L} \,,
\end{gather}
are determined from simulations in the usual way together with the two-point 
correlators defined in Eq.~\ref{e:ecor}.
The Euclidean endcap function in Eq.~\ref{e:02eef} yields the finite-volume smeared 
spectral function $\hat{\rho}_{\boldsymbol{p}_2,0}^{L,J,\epsilon}(q_1)$,
which approaches $\hat{\rho}_{\boldsymbol{p}_2,0}^{J,\epsilon}(q_1)$ as 
$L \rightarrow \infty$. 
Note that, in contrast to typical lattice calculations of three-point temporal 
correlation functions, the current insertion is placed at an earlier time than 
the two interpolators and only a 
single time separation is taken large.

The finite-volume reconstruction of 
$\hat{\rho}_{\boldsymbol{p}_2,0}^{L,J,\epsilon}(q_1)$ using Eq.~\ref{e:fvol}
may also be advantageous. At the on-shell point, the smeared spectral function decomposes into finite-volume energies and matrix elements as 
\begin{gather}\label{e:0to2rec} 
	\hat{\rho}_{\boldsymbol{p}_2,0}^{L,J,\epsilon}(-E(\boldsymbol{p}_1),-\boldsymbol{p}_1) =
	\sum_{n_1} \frac{i}{E(\boldsymbol{p}_2) + E(\boldsymbol{p}_1) - 
	E_{n_1}^L + i\epsilon}\, \sqrt{2 E(\boldsymbol p_2) L^3} \langle \boldsymbol{p}_2 | \hat{\phi}(-\boldsymbol{p}_1, 0) | n_1 \rangle_{L} \,\times \, \langle n_1 | \hat{J}(0) | 0 \rangle_{L}   \,.
\end{gather}
  The lowest 
finite-volume states contributing to the sum over $n_1$ resemble two-particle 
states as well as resonance-like states if present. The 
formalism for determining the energies and matrix elements of a few such 
low-lying states in lattice simulations is well developed. As an example, 
for the isovector time-like pion form factor in lattice QCD the matrix 
elements in Eq.~\ref{e:0to2rec} consist of 
$\langle \pi\pi | \hat{J}(0) | 0 \rangle_{L}$, where $\hat{J}$ is the 
isovector component of the electromagnetic current, 
and $\langle  \pi | \hat{\pi}(\boldsymbol{p}) | \pi\pi\rangle$, where 
$\hat{\pi}(\boldsymbol{p})$ is a single-pion interpolator. The former matrix elements are exactly those determined in existing lattice QCD calculations of the form factor~\cite{Feng:2014gba,Andersen:2018mau}, while the latter somewhat 
resemble the finite-volume matrix elements calculated 
 for the process $\pi+\gamma^* \rightarrow \pi\pi$  
(where $\gamma^*$ denotes a virtual photon) in 
Refs.~\cite{Briceno:2016kkp,Alexandrou:2018jbt} but with an arbitrary pion 
interpolator in place of the electromagnetic current.

\subsection{Exclusive two-to-two amplitudes}\label{s:2to2}

As a slightly more complicated example,  we next consider $\boldsymbol{p}_1 + 
\boldsymbol{p}_2 \rightarrow \boldsymbol{p}_3 + \boldsymbol{p}_4$  
without an external current. This process can be also 
treated below inelastic threshold using L\"{u}scher's finite-volume approach, 
but the method outlined here again yields the amplitude above 
arbitrary inelastic thresholds. The appropriate infinite-volume real-time endcap function is  
\begin{align}\label{e:2to2t}
	\widetilde{F}_{\boldsymbol{p}_4,\boldsymbol{p}_1}(q_3,q_2) & =  
	\int d^4 x_3 d^4x_2\, 
	{\rm e}^{-iq_3\cdot x_3} 
	{\rm e}^{-iq_2\cdot x_2}\, 
	\theta(t_3-t_2) \, \langle \boldsymbol{p}_4 | 
	\hat{\phi}(x_3)\,  
	\hat{\phi}(x_2) | \boldsymbol{p}_1 \rangle_{\rm c}  \,,
	\\\label{e:0to2lsz} 
&	= 
	\frac{Z^{1/2}(\boldsymbol{p}_3)}{2E(\boldsymbol{p}_3)}
	\frac{Z^{1/2}(\boldsymbol{p}_2)}{2E(\boldsymbol{p}_2)}
	\frac{i^2 (2\pi)^4 \delta^4(p_4 - q_3 - q_2 - p_1) }{ [ -q_3^0 - E(\boldsymbol{p}_3) + i\epsilon ] [q_2^0 - E(\boldsymbol{p}_2) + i\epsilon ]} 
	\, i{\cal M}_{\rm c}(p_4p_3|p_2p_1) \, + \cdots  \,, \\[5pt]\label{e:0to2sm}
& = 
	(2\pi)^4 \delta^4(p_4-q_3-q_2-p_1)\, \hat{\rho}^{\epsilon}_{\boldsymbol{p}_4\boldsymbol{p}_1}(q_2) \,,
\end{align}
where the LSZ reduction is applied in Eq.~\ref{e:0to2lsz}  and in Eq.~\ref{e:0to2sm} we have introduced the associated smeared spectral function, defined as
\begin{equation}
\hat{\rho}^{\epsilon}_{\boldsymbol{p}_4\boldsymbol{p}_1}(q_2) \equiv   \int_{0}^{\infty} \frac{dE_2}{\pi}
	\frac{i}{E(\boldsymbol{p}_1) + q_2^0 - E_2 + i\epsilon} 
	\rho_{\boldsymbol{p}_4\boldsymbol{p}_1}(E_2,\boldsymbol{p}_2)   \,.
\end{equation}

Following the now-established pattern, to obtain the finite-volume smeared spectral function from a 
lattice simulation we require
the Euclidean endcap function
\begin{align}\label{e:2to2}
	C_{\boldsymbol{p}_4\boldsymbol{p}_1}^{L}(\boldsymbol{p}_2,\tau_2) & =    2 \sqrt{E(\boldsymbol p_1) E(\boldsymbol p_4)} L^3
	\langle \boldsymbol{p}_4 | \hat{\phi}(0)\, {\rm e}^{\hat{H}\tau_2} 
	\hat{\varphi}(\boldsymbol{p}_2, 0) | \boldsymbol{p}_1 \rangle_{{\rm c},L} \,,
\\  
	&=   2 \sqrt{E(\boldsymbol p_1) E(\boldsymbol p_4)} L^3
	Z^{1/2}(\boldsymbol{p}_4)
	Z^{1/2}(\boldsymbol{p}_1) \lim_{{\tau_4, \atop \tau_2-\tau_1} \rightarrow \infty} 
	\frac{ C^{L}_{4{\rm pt}} \big ((-\boldsymbol{p}_4,\tau_4),(\boldsymbol{p}_2,\tau_2),(\boldsymbol{p}_1,\tau_1) \big )}{
		C^{L}_{2{\rm pt}}(\boldsymbol{p}_4,\tau_4)\,
		C^{L}_{2{\rm pt}}(\boldsymbol{p}_1,\tau_2-\tau_1)} \,,
\\ &	= \int_0^\infty \frac{dE_2}{\pi} \, {\rm e}^{E_2 \tau_2} \, \rho^{L}_{\boldsymbol{p}_4,\boldsymbol{p}_1}(E_2,\boldsymbol{p}_2) \,, 
\end{align}
where $\tau_2 < 0$, which is obtained from the four-point temporal correlation function  
\begin{gather}\label{e:4pt}
	C_{4{\rm pt}}^{L} \big ((-\boldsymbol{p}_4,\tau_4),(\boldsymbol{p}_2,\tau_2),(\boldsymbol{p}_1,\tau_1) \big )= 
	\langle 0 | \hat{\varphi}(-\boldsymbol{p}_4,0) \,{\rm e}^{-\hat{H}\tau_4}\, 
	\hat{\phi}(0)\, {\rm e}^{\hat{H}\tau_2} \,\hat{\varphi}(\boldsymbol{p}_2,0)\, 
	{\rm e}^{-\hat{H}(\tau_2 - \tau_1)} \,\hat{\varphi}(\boldsymbol{p}_1,0) | 0\rangle_{{\rm c},L}  \,,
\end{gather}
 and the two-point correlators defined 
in Eq.~\ref{e:ecor}. 
Four-point temporal correlation functions (with two time separations taken large) are not typically
calculated in lattice QCD simulations and are also required in 
Ref.~\cite{Hansen:2017mnd} for one-to-$X$ inclusive rates mediated by 
an external current.

The amplitude is obtained in the usual way by taking the limit 
ordered double limit of $L \rightarrow \infty$ followed by 
$\epsilon \rightarrow 0^+$ 
\begin{align}\label{e:2to2lim}
	i{\cal M}_{\rm c}(p_4p_3|p_2p_1)& =
	\lim_{\epsilon \rightarrow 0^+}    
	\lim_{L \rightarrow \infty}   \, 	
	i{\cal M}^{L,\epsilon}_{\rm c}(p_4p_3|p_2p_1) \,,
	\\\label{e:2to2lim2}
	i{\cal M}^{L,\epsilon}_{\rm c}(p_4p_3|p_2p_1) &\equiv 	\frac{2E(\boldsymbol{p}_3)}{Z^{1/2}(\boldsymbol{p}_3)}
	\frac{2E(\boldsymbol{p}_2)}{Z^{1/2}(\boldsymbol{p}_2)} \, \epsilon^2 \, \hat{\rho}^{L, \epsilon}_{\boldsymbol{p}_4\boldsymbol{p}_1}(
	E(\boldsymbol{p}_2),\boldsymbol{p}_2) \,,
\end{align}
of the smeared spectral function at the on-shell point multiplied by the 
amputation factor $\epsilon^2$. 
As in Sec.~\ref{s:0to2}, the finite-volume spectral function from the 
last line of Eq.~\ref{e:2to2} is used to define the finite-volume 
smeared spectral function 
$\hat{\rho}_{\boldsymbol{p}_4\boldsymbol{p}_1}^{L,\epsilon}(E(\boldsymbol{p}_2),\boldsymbol{p}_2)$ by convolution with the on-shell 
smearing kernel from Eq.~\ref{e:2ker}.

Finite-volume reconstruction of the smeared spectral function proceeds in a 
similar manner as in Sec.~\ref{s:0to2} by expressing it in terms of finite-volume energies and matrix elements 
\begin{gather}\label{e:2to2fv}
\hat{\rho}_{\boldsymbol{p}_4\boldsymbol{p}_1}^{L,\epsilon}(E(\boldsymbol{p}_2),\boldsymbol{p}_2) =   2 \sqrt{E(\boldsymbol p_1) E(\boldsymbol p_4)} L^3
	\sum_{n_1} \frac{i \, \langle \boldsymbol{p}_4 |  \hat{\phi}(0) | n_1 \rangle_{L} \, \times \, \langle n_1 | 
	\hat{\varphi}(\boldsymbol{p}_2,0) | \boldsymbol{p}_1 \rangle_{L} }{E(\boldsymbol{p}_1) + E(\boldsymbol{p}_2) - E_{n_1}^L + i\epsilon} \,-\, \cdots,
\end{gather}
where the ellipsis denotes the subtraction of disconnected diagrams. Determination of 
these finite-volume quantities for a few low-lying states may saturate the 
sum over $n_1$ or perhaps stabilize a reconstruction procedure. For the example of $\pi \pi \rightarrow \pi\pi$, 
the contribution from each state requires a 
finite-volume two-pion energy and two matrix elements of the form  
$\langle \pi | \hat{\pi} | \pi\pi \rangle_{L}$ discussed in 
Sec.~\ref{s:0to2} for $\gamma^* \rightarrow \pi\pi$. 

\bigskip 

As mentioned above,
zero-to-$X$ and one-to-$X$  inclusive rates mediated by an external current can already be treated using Ref.~\cite{Hansen:2017mnd}.
However, arbitrary inclusive transition rates can be calculated from our approach using the unitarity cut of Eq~\ref{e:ucut}.  
 As a minimal example, consider the 
two-to-$X$ inclusive rate  $\boldsymbol{p}_1 + 
\boldsymbol{p}_2 \rightarrow X$ without an external current. 
An example of such a purely hadronic inclusive rate is furnished by the total proton-proton cross section $pp \rightarrow X$ studied at the LHC by the TOTEM experiment~\cite{Antchev:2017dia}. 

In the case of two-to-two scattering, the unitarity cut 
simplifies to Eq.~\ref{e:2to2fv}, which becomes the optical theorem in the ordered double limit with $(p_4,p_3) = (p_1,p_2)$  
\begin{equation}\label{e:opt2}
	\sum_{\alpha} (2\pi)^4 
	\delta^4(p_{\alpha} -p_1-p_2) \left| {\cal M}(\alpha | p_1p_2) \right|^2 = 2{\rm Im} \,{\cal M}_{\rm c}(p_1p_2|p_1p_2)  
	= \frac{8E(\boldsymbol{p}_2)^2 }{Z(\boldsymbol{p}_2)} \, 
	\lim_{\epsilon\rightarrow 0^{+}} \lim_{L \rightarrow \infty} \,  
	\epsilon^2 \, {\rm Re} \, \hat{\rho}^{L,\epsilon}_{\boldsymbol{p}_1\boldsymbol{p}_1} (E(\boldsymbol{p}_2), \boldsymbol{p}_2)   \,,
\end{equation}
where Eq.~\ref{e:2to2lim2} is used in the last line. 
The unsmeared spectral function $\rho^{L}_{\boldsymbol{p}_1\boldsymbol{p}_1}(E,\boldsymbol{p}_2)$ with identical endcaps is positive definite, requiring only the solution of a single real inverse problem  
\begin{gather}\label{e:2toXsm}
	{\rm Re} \, \hat{\rho}^{L,\epsilon}_{\boldsymbol{p}_1\boldsymbol{p}_1} (E(\boldsymbol{p}_2), \boldsymbol{p}_2) = \int_0^{\infty} \frac{dE_2}{\pi} \frac{\epsilon}{
		(E(\boldsymbol{p}_1) + E(\boldsymbol{p}_2) - E_2)^2 + \epsilon^2} 
	\rho^{L}_{\boldsymbol{p}_1\boldsymbol{p}_1}(E_2,\boldsymbol{p}_2), 
\end{gather}
where we have taken the real part of the amputation factor given in 
Eq.~\ref{e:2ker}. As already mentioned in the introduction, this result has a 
strong the connection to the work of Ref.~\cite{Hansen:2017mnd}. 

\subsection {Three-to-three amplitudes}\label{s:3to3}

As a final example, consider the exclusive three-to-three process $\boldsymbol{p}_1 + \boldsymbol{p}_2 + \boldsymbol{p}_3 \rightarrow \boldsymbol{p}_4 + \boldsymbol{p}_5 + \boldsymbol{p}_6$ without an external current. This process is 
practically more complicated 
than the previous examples but conceptually straightforward. As is demonstrated
shortly, it requires the 
lattice calculation of six-point 
temporal correlation functions and the determination of a spectral function 
with three energy arguments. Given such difficulties, three-to-three amplitudes 
are likely a future application of the formalism. Nonetheless, it is instructive
to 
illustrate the relative theoretical simplicity of our 
three-to-three formalism compared to the significant complication 
incurred when treating three-to-three scattering in the finite-volume approach.
Three-to-three scattering processes are present in a number of phenomenological 
applications including the three body decay of the $N(1440)$ Roper resonance, the determination of  
three-nucleon forces, and study of the $\omega(782)$-resonance.  

As above, we begin by expressing the relevant infinite-volume real-time endcap function in terms of the scattering amplitude as well as the smeared spectral function  
\begin{align}
	\widetilde{F}_{\boldsymbol{p}_6\boldsymbol{p}_1}(q_5,q_4,q_3,q_2) & = 
	\int \prod_{j=2}^{5} \left\{ d^4x_j \, {\rm e}^{-iq_j \cdot x_j} \right\} 
	\, \theta(t_5-t_4) \theta(t_4-t_3) \theta(t_3-t_2)\,    
	\langle \boldsymbol{p}_6 | \hat{\phi}(x_5) \, \hat{\phi}(x_4) \, 
	\hat{\phi}(x_3) \, 
	\hat{\phi}(x_2) | \boldsymbol{p}_1 \rangle_{\rm c} \,,
	\\  
	& \hspace{-30pt} = \prod_{j=2}^{5} 
	\left\{ \frac{Z^{1/2}(\boldsymbol{p}_j)}{2E(\boldsymbol{p}_j)}\right\}
	\frac{i^4 	(2\pi)^4 \delta^4(p_6 - q_5 -q_4-q_3-q_2-p_1) \, i {\cal M}_{\rm c}(
	p_6p_5p_4|p_3p_2p_1)}{[ q_2^0 - E(\boldsymbol{p}_2) + i\epsilon ][q_3^0 - E(\boldsymbol{p}_3) + i\epsilon] [-q_4^0 - E(\boldsymbol{p}_4) + i\epsilon] [-q_5^0 - E(\boldsymbol{p}_5) + i\epsilon]} + \cdots \,,
	\\
	&  \hspace{-30pt}  = 	(2\pi)^4 \delta^4(p_6 - q_5 -q_4-q_3-q_2-p_1) \, \hat{\rho}_{\boldsymbol{p}_6\boldsymbol{p}_1}^{\epsilon}(q_5,q_3,q_2)  \,,
\end{align}
 which is now depends on three energy-momentum arguments. 
 In the last line we have introduced
\begin{equation}
\label{eq:3particleConv}
\hat{\rho}_{\boldsymbol{p}_6\boldsymbol{p}_1}^{\epsilon}(q_5,q_3,q_2) \equiv   \int_0^\infty \! \! \frac{d^3 E}{\pi^3} \,  \hat \delta^3_{\epsilon}(q_0, E)   \, 	\rho_{\boldsymbol{p}_6\boldsymbol{p}_1}\big ((E_5,-\boldsymbol{p}_5), (E_3,\boldsymbol{p}_3),(E_2,\boldsymbol{p}_2)\big) \,,
\end{equation}
 where
\begin{align}
 \int_0^\infty \! \! \frac{d^3 E}{ \pi^3} & \equiv \int_0^{\infty} \frac{dE_5}{\pi} \, 
	\frac{dE_3}{\pi}\, \frac{dE_2}{\pi}   \,,  \\
	\hat \delta^3_{\epsilon}(q_0, E) & \equiv \frac{i}{E(\boldsymbol{p}_6) - q_5^0 - E_5 +i\epsilon}  
	\frac{i}{E(\boldsymbol{p}_1) + q_2^0 + q_3^0 - E_3 + i\epsilon} 
	\frac{i}{E(\boldsymbol{p}_1) + q_2^0 - E_2 + i\epsilon}      \,.
\end{align}

The desired spectral function, $\hat{\rho}_{\boldsymbol{p}_6\boldsymbol{p}_1}^{\epsilon}$, is obtained from its finite-volume analog, ${\rho}^{L}_{\boldsymbol{p}_6\boldsymbol{p}_1}$, which is defined via its relation to the finite-volume Euclidean endcap function. The appropriate endcap function is  
\begin{align}\label{e:3to3e}
\begin{split}
	{C}^{L}_{\boldsymbol{p}_6\boldsymbol{p}_1} \big ((-\boldsymbol{p}_5,\tau_5), (\boldsymbol{p}_3,\tau_3),(\boldsymbol{p}_2,\tau_2) \big) & = 2\sqrt{E(\boldsymbol{p}_6) 
	E(\boldsymbol{p}_1)} L^3  
	\\  & \hspace{40pt} \times \langle \boldsymbol{p}_6 | \hat{\varphi}(-\boldsymbol{p}_5,0) {\rm e}^{-\hat{H}\tau_5} 
	\hat{\phi}(0)\, {\rm e}^{\hat{H}\tau_3} \hat{\varphi}(\boldsymbol{p}_3,0) \,
	{\rm e}^{-\hat{H}(\tau_3-\tau_2)}\hat{\varphi}(\boldsymbol{p}_2,0) | 
	\boldsymbol{p}_1 \rangle_{{\rm c},L} \,,
	\end{split}\\[5pt]  
	&  \hspace{-50pt} = 2\sqrt{E(\boldsymbol{p}_6) 
	E(\boldsymbol{p}_1)} L^3 \,
		Z^{1/2}(\boldsymbol{p}_6)
		Z^{1/2}(\boldsymbol{p}_1)
	\lim_{{\tau_6-\tau_5, \atop \tau_2-\tau_1 }\rightarrow \infty} \frac{{  C}_{6{\rm pt}}^{L}\big(
	(-\boldsymbol{p}_6,\tau_6),\dots,(\boldsymbol{p}_1,\tau_1) \big)}{
		C^{L}_{2{\rm pt}}(\boldsymbol{p}_6,\tau_6-\tau_5)\, 
		C^{L}_{2{\rm pt}}(\boldsymbol{p}_1,\tau_2-\tau_1)}  \,,
	\\ 
	&  \hspace{-50pt}  = \int_{0}^{\infty} \frac{dE_5}{\pi}\,\frac{dE_3}{\pi}\,\frac{dE_2}{\pi} \, {\rm e}^{-E_5\tau_5} \, 
	{\rm e}^{E_3\tau_3}\, {\rm e}^{-E_2(\tau_3-\tau_2)} \,  	{\rho}^{L}_{\boldsymbol{p}_6\boldsymbol{p}_1} \big ((E_5,-\boldsymbol{p}_5), (E_3,\boldsymbol{p}_3),(E_2,\boldsymbol{p}_2) \big )\,,
\end{align}
where $\tau_2 < \tau_3 <0 <\tau_5$, and the six-point temporal correlation function is 
\begin{multline}
	{C}^{L}_{6{\rm pt}} \big ((-\boldsymbol{p}_6,\tau_6), \dots, (\boldsymbol{p}_1,\tau_1) \big) = 
	\langle 0 | 
	\hat{\varphi}(-\boldsymbol{p}_6,0) {\rm e}^{-\hat{H}(\tau_6-\tau_5)} \,
	\hat{\varphi}(-\boldsymbol{p}_5,0) {\rm e}^{-\hat{H}\tau_5} \, \times
\\ 	
	\hat{\phi}(0) 
	{\rm e}^{\hat{H}\tau_3}	\hat{\varphi}(\boldsymbol{p}_3,0) 
	{\rm e}^{-\hat{H}(\tau_3-\tau_2)} \,
	\hat{\varphi}(\boldsymbol{p}_2,0) {\rm e}^{-\hat{H}(\tau_2-\tau_1)} \,
	\hat{\varphi}(\boldsymbol{p}_1,0) | 0 \rangle_{{\rm c}, L} \,.
\end{multline}

As with the previous examples, the finite-volume spectral function can also be expressed in terms of finite-volume matrix elements and energies 
\begin{multline}\label{e:3to3s}
\rho^{L}_{\boldsymbol{p}_6\boldsymbol{p}_1}((E_5,-\boldsymbol{p}_5), (E_3,\boldsymbol{p}_3),(E_2,\boldsymbol{p}_2)) = 2\sqrt{E(\boldsymbol{p}_6) 
	E(\boldsymbol{p}_1)} L^3 
	\sum_{n_5,n_3,n_2}\pi^3
	\, \delta(E_5 - E^L_{n_5})
	\, \delta(E_3 - E^L_{n_3})
	\, \delta(E_2 - E^L_{n_2}) \, \times 
	\\  
	\langle \boldsymbol{p}_6 | \hat{\varphi}(-\boldsymbol{p}_5,0) | 
	n_5 \rangle_L \, \langle n_5 | \hat{\phi}(0) |  
	n_3 \rangle_L \, \langle n_3 | \hat{\varphi}(\boldsymbol{p}_3,0) |
	n_2 \rangle_L \, \langle n_2 | \hat{\varphi}(\boldsymbol{p}_2,0) 
	| \boldsymbol{p}_1 \rangle_{L} \,,
\end{multline} 
so that the reconstruction procedure will aim to construct the smeared spectral function defined via Eq.~\ref{eq:3particleConv}.
The scattering amplitude is then related to the latter by
\begin{gather}\label{e:3to3a}
	i{\cal M}_{\rm c}(p_6p_5p_4 | 
	p_3p_2p_1) = \prod_{j=2}^{5} \left\{ 
	\frac{2E(\boldsymbol{p}_j)}{
		Z^{1/2}(\boldsymbol{p}_j)}\right\}\,  
	\lim_{\epsilon \rightarrow 0^+} 
	\lim_{L \rightarrow \infty} 
	\epsilon^4 \, \hat{\rho}^{L,\epsilon}_{\boldsymbol{p}_6\boldsymbol{p}_1}(
	(-E(\boldsymbol{p}_5),-\boldsymbol{p}_5),(E(\boldsymbol{p}_3),\boldsymbol{p}_3),
	(E(\boldsymbol{p}_2),\boldsymbol{p}_2)) \,.
\end{gather}

The exclusive connected three-to-three amplitude is therefore obtained in
manner completely analogous to the other examples, introducing no formal complications. However,
considerable practical complications are required, namely the 
evaluation of connected six-point temporal correlation functions and the 
solution of a three-dimensional inverse problem to determine the desired 
smeared spectral function.

The reconstruction of the three-dimensional smeared spectral function in 
Eq.~\ref{e:3to3s} is likely a difficult problem numerically. However, 
there are several techniques which may ameliorate 
this multi-dimesional inverse problem.
First, by treating individual terms in the sums over $n_5,n_3,n_2$ 
the smeared spectral function may be partially reconstructed by 
determining the corresponding finite-volume energies and matrix elements. For 
example, 
 reconstruction of a lattice QCD spectral function relevant for 
 the process $\pi \pi \pi \to \pi \pi \pi$ requires both two- and three-pion energies, and matrix elements of single-pion interpolators between single-, two-, and three-pion states. While the isolation of finite-volume 
 three-hadron states is somewhat beyound
 the current lattice QCD state-of-the art, it proceeds using establish methods. 

The unitarity cut approach of 
Eq.~\ref{e:ucut} may also be applied. To this end the 
smeared spectral function is expressed as 
	\begin{gather} 
	\hat{\rho}^{L,\epsilon}_{\boldsymbol{p}_6\boldsymbol{p}_1} \big (
	(-E(\boldsymbol{p}_5),-\boldsymbol{p}_5),(E(\boldsymbol{p}_3),\boldsymbol{p}_3),
	(E(\boldsymbol{p}_2),\boldsymbol{p}_2) \big)  = 
		 \sum_{n_3} \delta_{\boldsymbol{p}_{n_3},\boldsymbol{p}_{3} + \boldsymbol{p}_{2} + \boldsymbol{p}_1} 
		\frac{
		i\, \hat{\varrho}^{L,J,\epsilon}_{\boldsymbol{p}_{6},n_3} ( -E(\boldsymbol{p}_5),
	-\boldsymbol{p}_5)
		\,  
		\hat{\varrho}^{L,\epsilon}_{n_3,\boldsymbol{p}_1}(E(\boldsymbol{p}_2),
	\boldsymbol{p}_2)}{E(\boldsymbol{p}_1) + E(\boldsymbol{p}_2) + E(\boldsymbol{p}_3) - E^L_{n_3} + i\epsilon} - \cdots \,,
	\end{gather}
where the cut is performed at the central argument and the ellipsis denotes 
the explicit subtraction of disconnected terms.  
The smeared 
spectral functions with a single argument appearing in the numerator are
obtained from disconnected Euclidean endcap functions (without explicit subtractions) which have the arbitrary finite-volume state 
$|n_3\rangle_{L}$ as an endcap. 

 As a specific example, the 
smeared spectral function 
\begin{align}
	\hat{\varrho}^{L,\epsilon}_{\boldsymbol{p}_6,n_3} (-E(\boldsymbol{p}_5), 
	-\boldsymbol{p}_5) &= \sqrt{2E(\boldsymbol{p}_6)L^3}\,\sum_{n_5} \frac{i
	\langle \boldsymbol{p}_6 | \hat{\phi}(-\boldsymbol{p}_5,0) | n_5 \rangle_L \, \langle n_5 | \hat{\phi}(0) | n_3 \rangle_{L} }{
		E(\boldsymbol{p}_6) + E(\boldsymbol{p}_5) - E_{n_5}^L + i\epsilon}   \,,
\\  
	&= \int_{0}^{\infty} \frac{dE_5}{\pi} \frac{i}{E(\boldsymbol{p}_6) - 
	E(\boldsymbol{p}_5) - E_5 + i\epsilon} \varrho^{L}_{\boldsymbol{p}_6,n_3}(
	E_5, -\boldsymbol{p}_5) \,,
\end{align}
 is obtained from the Euclidean endcap function 
\begin{align}\label{e:3to3e2}
	{  C}_{\boldsymbol{p}_6,n_3}(-\boldsymbol{p}_5, \tau_5) &=
	\sqrt{2E(\boldsymbol{p}_6)L^3}\, \langle \boldsymbol{p}_6 | \hat{\phi}(-\boldsymbol{p}_5,\tau_5) \,
	\hat{\phi}(0) | n_3 \rangle_{L}  
	= \int_0^{\infty} \frac{dE_5}{\pi} {\rm e}^{-E_5\tau_5} \varrho^{L}_{\boldsymbol{p}_6,n_3}(
	E_5, -\boldsymbol{p}_5)  \,.
\end{align}
In the absence of any resonances-like states, the lowest states 
contributing to the sum over $n_3$ in Eq.~\ref{e:3to3s} are finite-volume 
three-particle states. For each these states two
smeared 
spectral functions of this kind are required. 
Since the endcap function in Eq.~\ref{e:3to3e2} can be obtained from the
same six- and two-point correlation functions as 
 the one from Eq.~\ref{e:3to3e}, the unitarity cut approach may aid in
 the solution of the three-dimensional 
 inverse problem.

\section{Perturbative test}\label{s:pt}
 
In this section, the approach outlined in Sec.~\ref{s:met} is tested perturbatively through next-to-leading order (NLO) in $\lambda \phi^4$-theory.  This theory is defined by the usual lagrangian density but with Euclidean signature
\begin{equation}
\mathcal L(x) = \frac12 Z \big (\partial_\mu \phi(x) \big) \big  (\partial_\mu \phi(x) \big) + \frac12 Z_m m^2 \phi(x)^2 + \frac{  Z_\lambda \lambda}{4!} \phi(x)^4 \,,
\end{equation}
where $Z, Z_m, Z_\lambda$ are adjusted so that $\langle 0 \vert \phi(0) \vert \boldsymbol{p} \rangle = 1$, $m$ is the physical pole mass, and the threshold scattering amplitude is given exactly by ${\cal M}_{\rm c}(p_4p_3|p_2p_1)|_{\rm thresh} = -\lambda$.
This perturbative test is performed for the exclusive two-to-two scattering process $\boldsymbol{p}_1 + \boldsymbol{p}_2 \rightarrow \boldsymbol{p}_3 + \boldsymbol{p}_4$ without an external current, for which the formulae of Sec.~\ref{s:2to2} are employed. The leading-order result is presented in Sec.~\ref{s:ptLO} and 
the $\mathcal O(\lambda^2)$-contribution to the imaginary part of the scattering amplitude in Sec.~\ref{s:ptNLO}. By calculating the Euclidean-time-dependent correlator and
the finite-volume spectral function of Eq.~\ref{e:2to2} to $\mathcal O(\lambda^2)$, the 
approach to the desired scattering amplitude in the ordered double limit 
 $ \lim_{\epsilon \rightarrow 0^+}  \lim_{L\rightarrow \infty}$ is also 
 examined in Sec.~\ref{s:ptNLO}.  

\subsection{Leading order}\label{s:ptLO} 

Before proceeding to the $\mathcal O(\lambda^2)$-calculation, the main points are illustrated at leading order. 
First, our novel application of the LSZ reduction procedure is examined using 
the  endcap function from Eq.~\ref{e:2to2t}.
To this 
end employ the LSZ procedure on the well-known leading-order expression for 
the time-ordered four-point function (with vacuum endcaps) to obtain
\begin{equation}\label{e:pt1}
	\langle \boldsymbol{p}_4 | T \left\{\hat{\phi}(0) \hat{\phi}(x_2)\right\} | \boldsymbol{p}_1 \rangle_{\rm c} 
=
	- i \lambda  \int d^4 x \, {\rm e}^{ i(   p_4   -    p_1)\cdot x} \int \frac{d^4k_2}{(2\pi)^4}\frac{d^4k_3}{(2\pi)^4}  \frac{ i^2 {\rm e}^{- i k_2 \cdot (x-x_2)}  {\rm e}^{ i k_3 \cdot x } }{[k_2^2 - m^2 + i \epsilon'][k_3^2 - m^2 + i \epsilon']}  + \mathcal O(\lambda^2)  \,,
\end{equation}
where we have amputated two propagators to project out the two endcap states. The pole prescription entering here is distinct from that introduced with the Fourier transform and is therefore denoted $\epsilon'$. The time ordering with $t_2 > 0$ is irrelevant, so we focus on Eq.~\ref{e:pt1} with $t_2 < 0$ where it coincides with $\theta(-t_2) \langle \boldsymbol{p}_4 | \hat{\phi}(0) \hat{\phi}(x_2) | \boldsymbol{p}_1 \rangle_{\rm c}$. 
Fourier transforming this object from $x_2$ to $(q_2^0 + i \epsilon, \boldsymbol p_2)$ and applying Eq.~\ref{e:2to2t} yields the desired spectral function 
\begin{align} 
\label{eq:ptLO1}
\hat{\rho}^{\epsilon}_{\boldsymbol{p}_4\boldsymbol{p}_1}(q_2)	& =  \frac{- i \lambda }{2 E(\boldsymbol p_3)}   %
\bigg [    \frac{ 1  }{ 2 E({\boldsymbol p_2} )   }  \frac{i}{q_2^0 - E(\boldsymbol p_2) + i \epsilon}  \frac{ i }{E(\boldsymbol p_4)  - E(\boldsymbol p_1) - E(\boldsymbol p_2) - E(\boldsymbol p_3) }       \\
	   & \hspace{50pt}  +   \frac{i}{[q_2^0 - E({\boldsymbol p_2})  + i \epsilon] [q_2^0 + E({\boldsymbol p_2})  + i \epsilon] }  \frac{i}{E(\boldsymbol p_1) + q_2^0 - E(\boldsymbol p_4)  - E(\boldsymbol p_3) + i \epsilon}  \label{eq:ptLO2}    \\
		& \hspace{50pt}	+ \frac{ 1 }{   2 E({\boldsymbol p_2})   }  \frac{i}{  q_2^0 + E(\boldsymbol p_2)  + i \epsilon}  \frac{i}{E(\boldsymbol p_1)  - E(\boldsymbol p_2)- E(\boldsymbol p_4) - E(\boldsymbol p_3) }          \bigg ]  + \mathcal O(\lambda^2)  \,. 
	\label{eq:ptLO3}
\end{align}
 While $\epsilon$ is 
kept finite, $\epsilon'$ is taken to zero immediately. 

The first term, Eq.~\ref{eq:ptLO1}, arises from integrating $t$ between $0$ 
and $\infty$. Since $t$ is the time coordinate of the leading-order 
$\lambda \phi^4$ insertion, this term corresponds to the interaction vertex 
located after the $\hat \phi(0)$ insertion. Thus it is not surprising that the 
corresponding (would-be) on-shell pole\footnote{The one-to-three process is kinematically forbidden here, but possible in theories with multiple species of scalar fields.} is accessed in the amputation procedure appropriate for 
three-to-one scattering and does not contribute to the two-to-two on-shell 
pole.  
The second and third terms, Eqs.~\ref{eq:ptLO2} and~\ref{eq:ptLO3}, arise from integrating $t$ from $-\infty$ to $0$. This generates contributions to both 
two-to-two and one-to-three, with the ambiguity arising since $t_2$ and $t$ are integrated over the same region. The middle term, Eq.~\ref{eq:ptLO2}, corresponds to the desired two-to-two scattering amplitude and is isolated using the amputation procedure of Sec.~\ref{s:2to2} 
\begin{equation}
 i {\cal M}_{\rm c}(p_4p_3|p_2p_1) = 	2E(\boldsymbol{p}_2)\, 2E(\boldsymbol{p}_3)\times \lim_{\epsilon\rightarrow 0^{+}}\epsilon^2 \hat{\rho}^{\epsilon}_{\boldsymbol{p}_4\boldsymbol{p}_1}(E(\boldsymbol{p}_2),\boldsymbol{p}_2)  
	= -i\lambda + {\cal O}(\lambda^2) \,,
\end{equation}
which gives the expected leading-order amplitude. 

The approach outlined in this work intends that 
$\hat{\rho}^{\epsilon}_{\boldsymbol{p}_4\boldsymbol{p}_1}(q_2)$
is calculated as the 
$L \rightarrow \infty$ limit of the corresponding finite-volume object. 
 Having verified the amputation procedure for the infinite-volume expressions, we now turn to the finite-volume endcap function which 
yields the finite-volume smeared spectral 
function $\hat{\rho}^{L,\epsilon}_{\boldsymbol{p}_4\boldsymbol{p}_1}(q_2)$. To compute the relevant Euclidean correlation function one can either consider the full object directly or else treat the finite-volume states that arise in a spectral decomposition. We find the latter approach more instructive as it differs from the infinite-volume analysis above. We thus begin with the result for the finite-volume matrix element

\begin{equation}
  \langle \boldsymbol k_1  \boldsymbol k_2 \vert \hat \phi(0) \vert \boldsymbol p \rangle_L   =  \frac{2 E(\boldsymbol p) L^3 [\delta_{\boldsymbol p, \boldsymbol k_1}+\delta_{\boldsymbol p, \boldsymbol k_2}]}{4 L^{9/2}  \sqrt{E(\boldsymbol k_2) E(\boldsymbol k_1) E(\boldsymbol p) }}  
	+  \frac{1}{4 L^{9/2}  \sqrt{E(\boldsymbol k_2) E(\boldsymbol k_1) E(\boldsymbol p) }}  \frac{\lambda}{(p - k_1 - k_2)^2 - m^2 } + \mathcal{O}(\lambda^2),
  \label{eq:lamphi4FVME}
\end{equation}
where the allowed momenta are discrete, satisfying $\boldsymbol p = 2 \pi \boldsymbol n/L$, and all states are normalized to unity. 
This result  determines the endcap function by inserting a complete set of states  %
\begin{align}\label{e:ptmr}
C^L_{\boldsymbol{p}_4\boldsymbol{p}_1}(\boldsymbol p_2,\tau_2) & =  \sqrt{2 E(\boldsymbol p_1) 2 E(\boldsymbol p_4)} L^3 \sum_n \langle \boldsymbol p_4 \vert \hat \phi(0) \vert n \rangle_L \langle n \vert \hat \varphi(\boldsymbol{p}_2,\tau_2) \vert \boldsymbol p_1 \rangle_L - C^{L,\text{disc}}_{\boldsymbol{p}_4\boldsymbol{p}_1}(\boldsymbol p_2,\tau_2)  \,, \end{align}
where the subtraction of the disconnected term is indicated explicitly.
Pulling out the $x_2$-dependence by acting on the neighboring states and employing Eq.~\ref{eq:lamphi4FVME} gives
\begin{multline}\label{e:pte1}
	C^L_{\boldsymbol{p}_4\boldsymbol{p}_1}(\boldsymbol{p}_2,\tau_2) =  \lambda    \frac{1}{2 E(\boldsymbol p_2) 2 E(\boldsymbol k_3)} \frac{ {\rm e}^{\tau_2 E(\boldsymbol p_2)  } -   {\rm e}^{\tau_2 [E(\boldsymbol k_3) + E(\boldsymbol p_4) - E(\boldsymbol p_1) ]} }{    E(\boldsymbol p_1) + E(\boldsymbol p_2) - E(\boldsymbol k_3)  - E(\boldsymbol p_4) } \\ 
	-    \lambda    \frac{1}{2 E(\boldsymbol p_2) 2 E(\boldsymbol k_3)}  \bigg [  \frac{ {\rm e}^{\tau_2 E(\boldsymbol p_2)  }}{  E(\boldsymbol p_1) + E(\boldsymbol p_2) - E(\boldsymbol p_4) + E(\boldsymbol k_3)}  +    \frac{ {\rm e}^{\tau_2 [E(\boldsymbol k_3) + E(\boldsymbol p_4) - E(\boldsymbol p_1) ]   }}{ E(\boldsymbol k_3) + E(\boldsymbol p_4) - E(\boldsymbol p_1) + E(\boldsymbol p_2)} \bigg ] + \mathcal O(\lambda^2)
 \,.
\end{multline}
where the  Kronecker-$\delta$'s  in Eq.~\ref{eq:lamphi4FVME} select a single term from the sum over finite-volume two-particle states in Eq.~\ref{e:ptmr}. Here we have also introduced the shorthand $\boldsymbol k_3 \equiv \boldsymbol p_1 + \boldsymbol p_2 - \boldsymbol p_4$.
Note the resemblance between Eqs.~\ref{e:pte1} 
and~\ref{eq:ptLO1}-\ref{eq:ptLO3}. Apart from restricting the momenta to those allowed in the finite volume, there are no finite-volume effects in the endcap function at  ${\cal O}(\lambda)$. 
Recall that $C^L_{\boldsymbol{p}_4\boldsymbol{p}_1}(\boldsymbol{p}_2,\tau_2)$ is directly evaluated in a lattice calculation by taking the asymptotic limits of the ratio in Eq.~\ref{e:2to2}.

As expected, Eq.~\ref{e:pte1} is a smooth function of its momenta 
$\boldsymbol p_1,\boldsymbol p_2,\boldsymbol p_4 $. If these are tuned such 
that $  E(\boldsymbol p_1) + E(\boldsymbol p_2) - E(\boldsymbol k_3)  - E(\boldsymbol p_4) = 0$ then a cancellation arises between numerator and 
denominator leading to a term linear in $\tau_2$ 
\begin{equation}
\label{eq:oncorr}
	 C^L_{\boldsymbol{p}_4\boldsymbol{p}_1}(\boldsymbol{p}_2,\tau_2) =    \lambda    \frac{ {\rm e}^{\tau_2 E(\boldsymbol p_2)  }}{2 E(\boldsymbol p_2) 2 E(\boldsymbol k_3)}  \bigg [   \tau_2
-   \frac{ 1}{  2 E(\boldsymbol k_3)}  -    \frac{ 1 }{2 E(\boldsymbol p_2)}   \bigg ] + \mathcal O(\lambda^2) \qquad \text{on-shell\ momenta} \,.
\end{equation}
  The next step is to
employ one of the spectral reconstruction methods discussed in Sec.~\ref{s:calc} to effect  the substitution ${\rm e}^{\tau_2 E} \longrightarrow  i /(q_2^0 - E + i \epsilon) $ which converts the Euclidean endcap function to a smeared spectral function. Applying this step to Eq.~\ref{e:pte1} yields
\begin{align}
\begin{split}
	\hat{\rho}^{L,\epsilon}_{\boldsymbol{p}_4\boldsymbol{p}_1}(q_2) & =      \frac{\lambda}{2 E(\boldsymbol p_2) 2 E(\boldsymbol k_3)} \frac{ 1}{    E(\boldsymbol p_1) + E(\boldsymbol p_2) - E(\boldsymbol k_3)  - E(\boldsymbol p_4) }\\
	& \hspace{140pt} \times  \bigg [ \frac{i}{q_2^0 - E(\boldsymbol p_2) + i \epsilon} -    \frac{i}{  q_2^0 -  E(\boldsymbol k_3) - E(\boldsymbol p_4) + E(\boldsymbol p_1)  + i \epsilon }    \bigg ] + \cdots \,, 
	\end{split}  \\
& =    \frac{- i \lambda}{2 E(\boldsymbol p_2) 2 E(\boldsymbol k_3)}     \frac{i}{q_2^0 - E(\boldsymbol p_2) + i \epsilon}     \frac{i}{  q_2^0 -  E(\boldsymbol k_3) - E(\boldsymbol p_4) + E(\boldsymbol p_1)  + i \epsilon }  + \cdots  \,, %
\end{align}
where all terms not proportional to the desired double pole are denoted by the ellipses. Note that the first and second lines are exactly equal, in particular 
no additional terms have been absorbed into the ellipses. We thus see that the 
$1/\epsilon^2$ behavior is achieved by the difference of exponentials leading 
to a difference in poles that is identically equal to a double pole. 
The single complex pole  introduced by the smearing kernel therefore yields a product of two poles each regulated by the same $\epsilon$, as expected 
from the LSZ reduction.  

Alternatively the calculation can be performed with the momenta tuned to satisfy exact energy conservation. Then the correlator is that of Eq.~\ref{eq:oncorr}. Note that any linear mapping which replaces ${\rm e}^{\tau_2 E} \longrightarrow  i /(q_2^0 - E + i \epsilon) $ will also give $\tau_2 {\rm e}^{\tau_2 E} 
\longrightarrow  i /(q_2^0 - E + i \epsilon)^2 $. In this case Eq.~\ref{eq:oncorr} is taken to
\begin{equation}
	\hat{\rho}^{L,\epsilon}_{\boldsymbol{p}_4\boldsymbol{p}_1}(q_2) =  -i\lambda    \frac{ 1}{2 E(\boldsymbol p_2) 2 E(\boldsymbol k_3)}  \frac{i}{q_2^0 - E(\boldsymbol p_2) + i \epsilon} \bigg [ \frac{i}{q_2^0 - E(\boldsymbol p_2) + i \epsilon}
-   \frac{ i}{  2 E(\boldsymbol k_3)}  -    \frac{ i }{2 E(\boldsymbol p_2)}   \bigg ] + \mathcal O(\lambda^2)   \,,
\end{equation}
which also yields the required result after amputation. 

\subsection{Next-to-leading order}\label{s:ptNLO}

A limitation of the leading-order illustration presented above is that the 
$\mathcal O(\lambda)$-contribution to $\hat{\rho}^{\epsilon}_{\boldsymbol{p}_4\boldsymbol{p}_1}(q_2)$ does not contain a sum over all finite-volume 
two-particle states. This sum, which is a crucial aspect of non-perturbative 
spectral functions, first appears in the NLO calculation presented in this section. 
 To this end, we consider only the imaginary part of the scattering 
amplitude 
\begin{gather}\label{e:NLO1}
	\text{Im} \, \mathcal M_{\rm c}(p_4p_3|p_2p_1) =     \frac{\lambda^2  \sqrt{s/4 - m^2} }{16 \pi \sqrt{s}} + \mathcal O(\lambda^3)   \,, 
\end{gather}
where $s = E_{\rm cm}^2 = (E(\boldsymbol{p}_1) + E(\boldsymbol{p}_2))^2 - (\boldsymbol{p}_1 + \boldsymbol{p}_2)^2$ is the usual Mandelstam variable. 
This restriction allows us to demonstrate the role of summing over states without computing overly complicated finite-$\epsilon$ expressions.  

Calculating the finite-volume endcap function from Eq.~\ref{e:ptmr} to NLO gives  
\begin{align}\label{e:eefNLO}
C_{\boldsymbol{p}_4\boldsymbol{p}_1}(\boldsymbol{p}_2,\tau_2) & = 
	- \frac{\lambda^2}{2}   \ %
	   \frac{1}{L^3} \sum_{\boldsymbol k'}    \bigg [ \mathcal T_a(\tau_2, k')    + \mathcal T_b(\tau_2, k')    + \mathcal T_c(\tau_2, k')       \bigg ] \,,
\end{align}
where
\begin{align}
\mathcal T_a(\tau_2, k') & = -  \frac{   {\rm e}^{   k_2^0  \tau_2  }     }{ 2 E(\boldsymbol p_2)  [(p_1 + k_2 - p_4)^2 - m^2 ]}       \frac{1}{(2 E(\boldsymbol k') )^2   [ k_2^0+E(\boldsymbol p_1) - 2 E(\boldsymbol k')   ]}  \bigg \vert_{k_2^0 = E(\boldsymbol p_2)} \,, \\
\mathcal T_b(\tau_2, k') & = -     \frac{ 1      }{  [k_2^2 - m^2 ]  [(p_1 + k_2 - p_4)^2 - m^2 ]}    \frac{    {\rm e}^{    k_2^0    \tau_2 } }{(2 E(\boldsymbol k') )^2  }   \bigg \vert_{k_2^0 = 2 E(\boldsymbol k')  - E(\boldsymbol p_1)  }  \,, \\
	\mathcal T_c(\tau_2, k') & =  - \frac{   {\rm e}^{ k_2^0 \tau_2  }     }{2 E(\boldsymbol k_3)[k_2^2 - m^2 ]  }         \frac{1}{(2 E(\boldsymbol k') )^2    [ k_2^0 + E(\boldsymbol p_1) - 2 E(\boldsymbol k')   ]}  \bigg \vert_{k_2^0  = E(\boldsymbol p_1 + \boldsymbol p_2 - \boldsymbol p_4) + E(\boldsymbol p_4) - E(\boldsymbol p_1) }\,. \\
\end{align}

The usual spectral reconstruction procedure is now applied to effect the 
replacement of decaying exponentials with pole factors. For simplicity, 
consider the zero total three-momentum frame $\boldsymbol{p}_1 + 
\boldsymbol{p}_2 = \boldsymbol{p}_3 + \boldsymbol{p}_4 = 0$. Furthermore, 
we immediately set $q_2^0 = E(\boldsymbol p_2)$ so that 
the smeared spectral function becomes
	\begin{equation}\label{e:ssNLO}
		\hat{\rho}_{\boldsymbol{p}_4\boldsymbol{p}_1}^{L,\epsilon}(E(\boldsymbol{p}_2),\boldsymbol{p}_2)=  \frac{i}{  E^{2}_{\rm cm} \epsilon^2}  \frac{\lambda^2}{2}   \frac{1}{L^3} \sum_{\boldsymbol k'}^{\Lambda} \frac{1}{(2 E(\boldsymbol k'))^2}  \frac{ 1}{     ( E_{\rm cm} -2 E(\boldsymbol k')+i \epsilon)} \bigg [ 1  - \frac{ \epsilon^2}{4 E(\boldsymbol k')^{2}}   -  \frac{ \epsilon     (\epsilon+2 i E(\boldsymbol k')) }{E_{\rm cm} E(\boldsymbol  k') } \bigg ] \,.
\end{equation}
The familiar product of the $\epsilon^2$-pole and energy denominators in the 
first factor is removed by the amputation procedure to yield our estimator for the desired amplitude  
	\begin{equation}\label{e:estNLO}
		{\rm Im}\, {\cal M}_{\rm c}^{L,\epsilon}(p_4p_3|p_2p_1) =   \frac{\lambda^2}{2}   \frac{1}{L^3} \sum_{\boldsymbol k'}^{\Lambda} \frac{1}{(2 E(\boldsymbol k'))^2} \,  \text{Im} \left\{ \frac{ 1}{     ( E_{\rm cm} -2 E(\boldsymbol k')+i \epsilon)} \bigg [ 1  - \frac{ \epsilon^2}{4 E(\boldsymbol k')^{2}}   -  \frac{ \epsilon     (\epsilon+2 i E(\boldsymbol k')) }{E_{\rm cm} E(\boldsymbol  k') } \bigg ]\right\} \,.
\end{equation}
 In Eqs.~\ref{e:ssNLO} and \ref{e:estNLO} we have included a cutoff $\Lambda$ on the sum indicating that the latter must be regulated. The divergence here is non-standard, arising from the replacement of the decaying exponential with a pole factor that has a slower large-energy fall-off. In a numerical lattice calculation this manifests as discretization effects, the detailed investigation of 
 which goes beyond the scope of this work.  

 In the present application, this divergence only arises for nonzero $\epsilon$, and Eq.~\ref{e:estNLO} approaches ${\rm Im}\, {\cal M}_{\rm c}(p_4p_3|p_2p_1)$ in the appropriate ordered double limit  
	\begin{align}\label{e:dlNLO}
	\lim_{\epsilon \rightarrow 0^+}	
		\lim_{L\rightarrow \infty}  
			{\rm Im}\, {\cal M}_{\rm c}^{L,\epsilon}(p_4p_3|p_2p_1)
		 &=   \frac{\lambda^2}{2}  \int \frac{d^3 \boldsymbol k'}{(2 \pi)^3}  \frac{1}{(2 E(\boldsymbol k'))^2}  \text{Im} \frac{ 1}{     ( E_{\rm cm} -2 E(\boldsymbol k')+i 0^+)}  
		\\\nonumber  &=  \frac{\lambda^2  \sqrt{E^2_{\rm cm}/4 - m^2} }{16 \pi  E_{\rm cm}}    \,,
\end{align}
independently of $\Lambda$ as expected. 

This NLO example illustrates all the key features of the LSZ approach. The finite-volume smeared spectral function of Eq.~\ref{e:ssNLO} is 
accessible from Euclidean lattice simulations and manifestly contains a sum 
over all finite-volume states. However, at finite $\epsilon$ the pole factor in
the sum over $\boldsymbol{k}'$ damps out energies increasingly different from $E_{\rm cm}$ so that those states with $2E(\boldsymbol{k}') \approx E_{\rm cm}$ 
are most important. The spectral reconstruction algorithm which replaces the 
Euclidean decaying exponentials in Eq.~\ref{e:eefNLO} with the $i\epsilon$-type 
smearing kernels of Eq.~\ref{e:ssNLO} is crucial. In addition to enabling a well-defined infinite-volume limit, it effectively performs an analytic 
continuation from Euclidean to Minkowski time and selects the correct 
time-ordering. While numerically difficult 
in practice, this step can be performed exactly here. 
  It is also clear that the order of the double limit in Eq.~\ref{e:dlNLO} is crucial  
to recover the correct amplitude. 

\begin{figure}
\includegraphics[width=\textwidth]{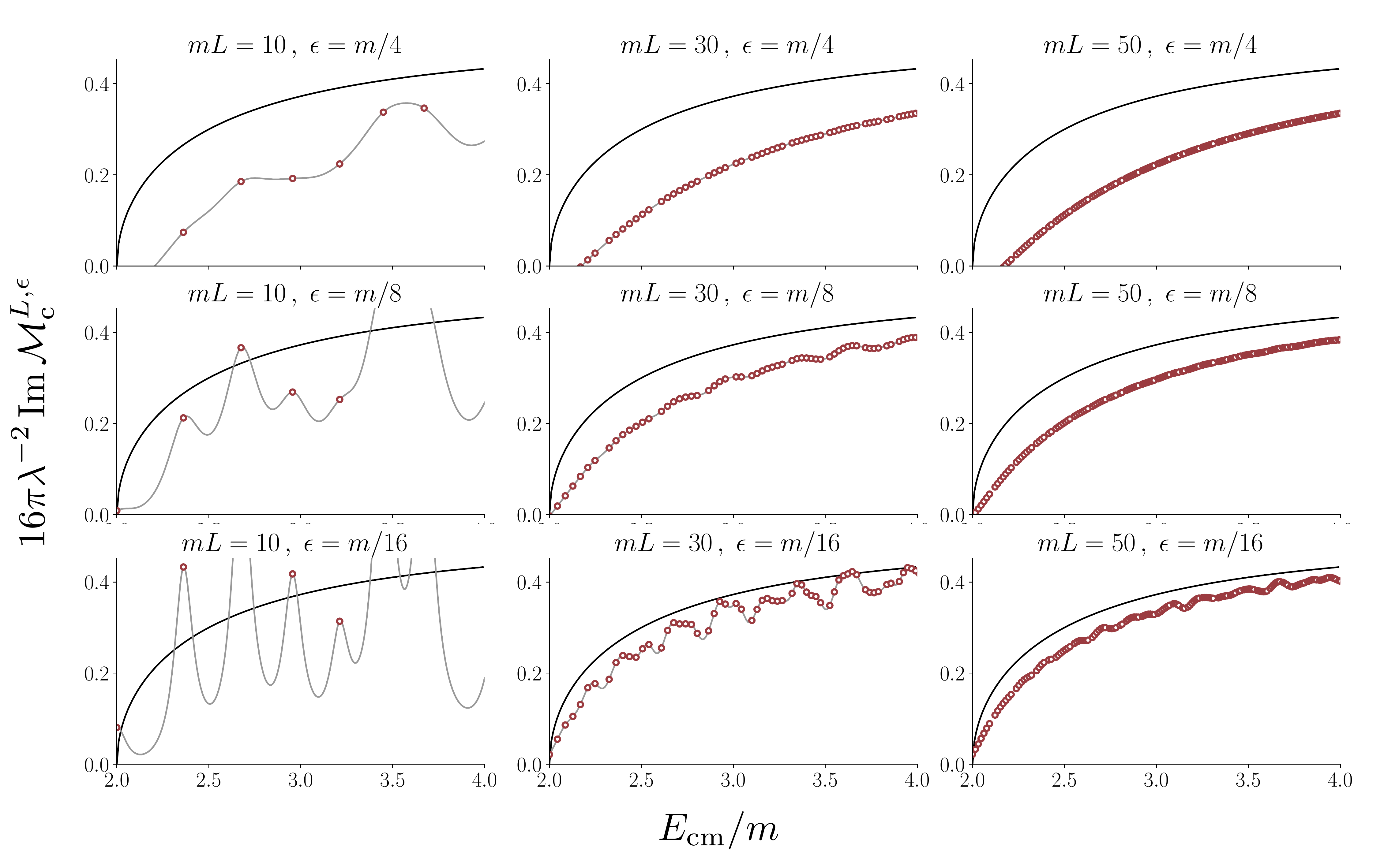}
	\caption{\label{f:lim}The estimator ${\rm Im}\, {\cal M}_{\rm c}^{L,\epsilon}$ from Eq.~\ref{e:estNLO} (with $\Lambda = 30m$) for various values of $mL$ and $\epsilon/m$. 
	The volume increases from left to right and $\epsilon$ decreases from top to bottom. The desired ordered double limit is therefore approached by first extrapolating to the right and then down. Taking the limit in the incorrect order by
	decreasing $\epsilon/m$ at fixed $mL$ (going down a column) reveals 
	contributions from individual finite-volume states, the values of which are indicated by the points.}
\end{figure}

The estimator in Eq.~\ref{e:estNLO} also provides some first 
indications of how the ordered double limit in $\epsilon$ and $L$ is approached.
The smearing kernel width $\epsilon$ adds a new scale to the usual infrared
hierarchy so that
$1/L \ll \epsilon \ll m$ is required to enter the asymptotic regime. Exploration of these limits 
is illustrated in Fig.~\ref{f:lim} where the estimator from 
Eq.~\ref{e:estNLO} is compared with the exact NLO result in Eq.~\ref{e:NLO1}.
Among the nine different choices for $\epsilon L$ and $\epsilon / m$ shown 
there, the desired ordered double limit is achieved by first extrapolating to 
the right in each row and then down the columns. 
From Fig.~\ref{f:lim} the necessity of the limit ordering is also apparent. 
Increasing the resolution by decreasing $\epsilon/m$ at fixed $mL$ (proceeding down a single column) 
reveals the contributions from individual finite-volume states. These 
individual contributions are decreasingly evident as $m L$ is increased at 
fixed $\epsilon$. 

\begin{figure}
\includegraphics[width=\textwidth]{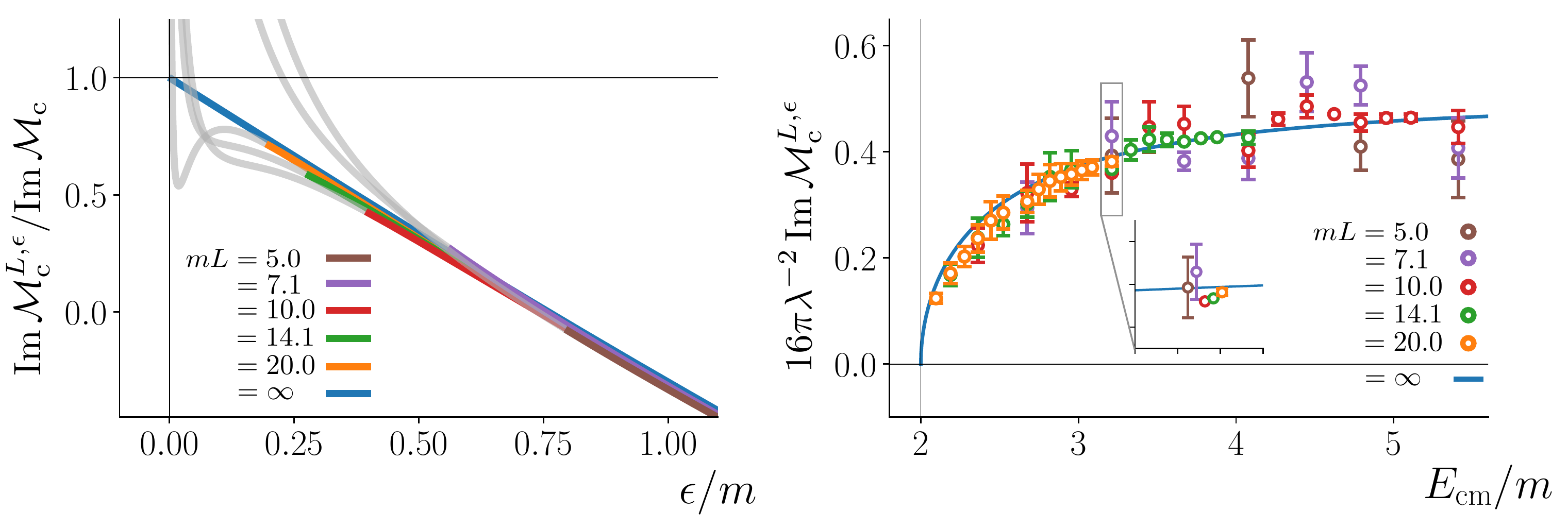} 
	\caption{\label{f:ext} Extrapolations to $\epsilon = 0$ at fixed $L$, keeping $\epsilon > 4/L$. {\bf Left}: The estimator from Eq.~\ref{e:estNLO} for a set of volumes increasing in steps of $\sqrt{2}$ so that $E_{\text{cm}}$ is 
	kept constant in physical units
	by adjusting the mode number $\boldsymbol n$, which is set to
	$\boldsymbol{n}^2 = 1$ at $mL = 5$. The vertical axis is normalized
	to the correct infinite-volume, zero-$\epsilon$ value. For each $L$ the point $\epsilon = 4/L$ is indicated by the transition from colored to gray. As the 
	volume increases larger regions of the curve approach the $mL=\infty$ 
	behavior, which is also shown. {\bf Right}: Set of extrapolations $\epsilon \to 0$ at fixed $L$, as explained in the text. The inset shows the five points corresponding to the left panel. Here the central values are given by averaging the linear and quadratic fits in $\epsilon$ with a systematic uncertainty  taken
	from their difference.}
\end{figure}

 Fig.~\ref{f:lim} also gives the first indications of values for which the 
estimator in Eq.~\ref{e:estNLO} is near the desired amplitude, suggesting 
that $mL \gtrsim 50$ is required for somewhat accurate results. This rather 
stringent guideline is relaxed significantly if extrapolations of $\epsilon/m \rightarrow 0$ are performed over a range of values at fixed $L$ which 
maintain $\epsilon L \gtrsim 4$. Such an extrapolation is illustrated in the left panel of
Fig.~\ref{f:ext}, where the estimator from Eq.~\ref{e:estNLO} is an apparently 
linear function of $\epsilon/m$ over the appropriate range. In the right panel of Fig.~\ref{f:ext} we show the result of extrapolations for a wide range of volumes and energies. For a particular value of $L$ and $E_{\text{cm}}= 2 \sqrt{m^2 +(2 \pi \boldsymbol n/L)^2}$, the extrapolation is performed by comparing linear and quadratic fits in $\epsilon$ over the range $[4/L, \epsilon_{\text{max}}(p, L)]$. Here we have introduced $p^2 = E_{\text{cm}}^2/4 - m^2$ and have set
\begin{equation}
 \epsilon_{\text{max}}(p, L) =   \begin{cases} 
      5/L & 5/L < p \\
    p & 5/L \leq p < m \\
      m & m \leq p  
   \end{cases}\,.
\end{equation}
This choice is motivated by the observation that the convergence is set by the  branch point which is a distance $p$ from the desired on-shell pole. Rather than always taking $p$ as the upper limit, the maximum is truncated on either side by additional considerations: \emph{(i)} the upper range should be well separated from $4/L$ and \emph{(ii)} the particle mass sets a second scale that should not be exceeded. We stress that this approach is only a first step and 
a more detailed analysis is required in numerical applications.

\section{Conclusions and outlook}\label{s:concl}

This work details a novel approach for determining scattering amplitudes from finite-volume Euclidean lattice field theory simulations.  
It is based on a relationship derived using the LSZ formalism between  finite-volume spectral functions and arbitrary real-time infinite-volume scattering amplitudes.  

Since the  
spectral function carries no information about the metric signature, it serves as a natural bridge between Euclidean and Minkowski time. The extraction of 
complex-valued  scattering amplitudes requires the convolution of the spectral
function with a particular complex smearing kernel which enforces the correct time ordering by implementing the $i\epsilon$-prescription. The amplitude is 
then recovered in three steps: {\em (i)} The limit $L \to \infty$ is saturated at fixed kernel-width $\epsilon$. {\em (ii)} A modified LSZ amputation is performed by multiplying with $\epsilon$ and dividing out operator overlaps. {\em (iii)} The limit $\epsilon \to 0^+$ is taken on the remaining, amputated object.     
The somewhat delicate interplay between the ordered double limit $ \lim_{\epsilon \rightarrow 0^+}  \lim_{L\rightarrow \infty}$, 
the analytic continuation between Euclidean and Minkowski time, the $i\epsilon$-prescription introduced by the smearing kernel, and the LSZ amputation procedure 
is illustrated for several examples in Sec.~\ref{s:appl} and perturbatively for two-to-two scattering 
in Sec.~\ref{s:pt}.

Although the formalism underpinning this approach is conceptually straightforward, its implementation in lattice QCD simulations consists of several practical 
complications. First, connected higher $n$-point temporal correlation functions that are not typically calculated in lattice QCD are required, 
although zero-to-two processes such as the timelike pion form factor simply require conventional three-point correlators. In order to specify the momenta of each scatterer, $n-1$ of the hadron interpolators 
in these higher-order $n$-point functions must be projected onto a definite spatial momentum. The efficient evaluation of each Wick contraction, the number of 
which proliferates rapidly with $n$, likely presents a challenge. In addition to difficulties in calculating the $n$-point functions, the endcap functions from Eq.~\ref{e:eef}
are obtained by taking the 
asymptotic limit of time separations required to isolate the endcap states. The ubiquitous signal-to-noise problems plaguing standard lattice 
QCD calculations are therefore also relevant here.  

Another practical difficulty concerns the determination of smeared spectral functions from Euclidean correlators.  
This effective replacement of decaying exponentials with $i\epsilon$-prescription pole factors is accomplished by 
the solution of an inverse problem. 
One possible strategy is a modification of the 
Backus-Gilbert approach, along the lines recently presented in 
Ref.~\cite{Hansen:2019idp}, in which the desired smearing kernel is treated as one of the inputs.  Although the solution of such inverse problems is 
 challenging, some potentially helpful prospects are  
discussed in Sec.~\ref{s:calc}. These include 
the separate determinations of the real and imaginary parts of the spectral 
function in Eq.~\ref{e:reim}, the determination of contributions from 
individual finite-volume states in Eq.~\ref{e:fvol}, the freedom of operator choice afforded by our amputation procedure, the prospect of using known properties of the $\epsilon$-dependence, and application of 
the unitarity cut formula of Eq.~\ref{e:ucut}.

The final practical difficulty foreseen for this approach is that it requires larger volumes than those conventionally applied in lattice QCD simulations. While 
$m_{\pi}L \gtrsim 4$ is generally sufficient to suppress undesirable finite-volume effects in current calculations, the parameter $\epsilon$ introduces another infrared scale into the usual hierarchy. Specifically one must achieve $1/L \ll \epsilon \ll m_{\pi}$.
 Some indication of suitable values for $\epsilon L$ and $\epsilon/m_{\pi}$ is given in 
Sec.~\ref{s:pt}, which suggests that $m_{\pi}L \approx 10-20$ may be 
sufficient. 

 For clarity, the formalism has been presented here for a single species of real scalar field.  The straightforward generalization to complex arbitrary-spin fields with internal degrees of freedom is accomplished by replacing $\hat{\phi}(x)$ with the required interpolator and applying the standard modification to the LSZ formalism.
	Multiple species are also easily handled. To unambiguously select a 
	desired in- or out-state, each individual field is placed on the mass shell of the desired particle. In contrast to the finite-volume L\"{u}scher approach, 
	additional spins and dynamical coupled scattering channels present no 
	real difficulties here since the LSZ reduction is concerned with 
	single-particle interpolators each of which
	are placed on shell individually. Particular initial and final states can therefore be unambiguously selected.  

	Another considerable advantage of this approach compared to L\"{u}scher-type methods is its validity above arbitrary inelastic thresholds. This potentially 
	extends the energy range for lattice QCD calculations of scattering amplitudes to encompass a number of interesting phenomenological applications. In particular, it provides
	a path by which many excited hadron resonances may be studied rigorously using lattice methods for the first time.   
	Overall the outlook for the methods introduced here depends on the degree to which the practical difficulties discussed above are controlled. 
	It is our view that the 
	simplicity of the LSZ approach to lattice QCD calculations of scattering amplitudes justifies significant 
	investment in overcoming these obstacles.

\vspace{4mm} 
\noindent 
{\bf Acknowledgements}:
We acknowledge helpful conversations with M.~Bruno, M.~Della Morte, 
 M.~L\"{u}scher, H.~Meyer, and N.~Tantalo. We also gratefully thank the Galileo Galilei Institute for Theoretical Physics 
and the Mainz Institute for Theoretical Physics for hospitality and support during completion of this work. Finally, we thank A. Francis and J. Ghiglieri, who, as the organizers of the CERN TH Institute ``From Euclidean spectral densities to real-time physics'', provided an environment of insightful talks and discussion at the latter stages of the project.

\appendix

\section{LSZ reduction for endcap functions}\label{a:lsz}  

Here the LSZ reduction formula from Eq.~\ref{e:tlsz} is proven for a general 
$m$-to-$n$ exclusive transition mediated by a local external current $\hat{J}(x)$. The less formal but more illustrative approach of 
Ref.~\cite{Sterman:1994ce} is employed. All of what follows is in 
infinite-volume Minkowski space.  

Consider the endcap function defined in Eq.~\ref{e:toef}.
We begin by reducing the rightmost field to isolate a two-particle in-state on 
the right endcap. Insert a complete set of two-particle in-states 
and examine 
the asymptotic contribution to the integral from the region $t_2 < -t_{\rm c} < t_3$
with $t_{\rm c} > 0$ taken arbitrarily large. This gives 
\begin{equation}
\begin{split}\label{e:lpr1}
	\widetilde{F}^{J}_{\boldsymbol{p}_{m+n}\boldsymbol{p}_1}(q_{r+1},\dots,q_2) & = 
	\int\prod_{j=3}^{r+1} \left\{ d^4x_j\,{\rm e}^{-iq_j\cdot x_j} \right\}
	\, \theta(t_{r+1}-t_r) \cdots \theta(t_4-t_3)  \int d^3 \boldsymbol{x}_2 \int_{-\infty}^{-t_{\rm c}} dt_2 \, {\rm e}^{-iq_2\cdot x_2}  \,   
\\ & \hspace{-80pt} \times
	 \int  \! d\Gamma(\boldsymbol{k}_1) 
	d\Gamma(\boldsymbol{k}_2) \  
	\langle \boldsymbol{p}_{m+n} | \hat{\phi}(x_{r+1}) \cdots \hat{\phi}(x_{m+1}) 
	\, \hat{J}(0) \hat{\phi}(x_m) \cdots \hat{\phi}(x_3) | \boldsymbol{k}_1
	\boldsymbol{k}_2 \rangle_{\rm in,c}\ {_{\rm in} \! \langle} \boldsymbol{k}_1 \boldsymbol{k}_2 | \hat{\phi}(x_2) |
	\boldsymbol{p}_1 \rangle + \cdots \,,  
\end{split}
\end{equation}
 where
the ellipsis denotes contributions from ($k\ne2$)-particle states and the remaining $t_2$-integration. In inserting the complete set of two-particle states the 
symmetry factor $1/2!$ has been neglected since it will be cancelled by a 
subsequent multiplicity.

Since we are interested only in the single-particle pole (in $q_2^0)$ of 
Eq.~\ref{e:lpr1} which occurs due to the $t_2 \rightarrow -\infty$
contribution to the integral, the integrand may be replaced by its value in 
this limit. For this limiting value, the asymptotic formalism of Haag and 
Ruelle~\cite{Haag,Ruelle} may be applied 
\begin{align}
	\lim_{t_2 \rightarrow -\infty} {_{\rm in} \langle } \boldsymbol{k}_1 
	\boldsymbol{k}_2 | \hat{\phi}(x_2) | \boldsymbol{p}_1\rangle &= 
	\langle \boldsymbol{k}_1 | \boldsymbol{p}_1 \rangle \, \times \, 
	\langle \boldsymbol{k}_2 | \hat{\phi}(x_2) | 0 \rangle  \,,
\\
	&= 
	(2\pi)^3 \, 2E(\boldsymbol{p}_1) \, \delta^3(\boldsymbol{k}_1 - \boldsymbol{p}_1 )\, {\rm e}^{ik_2\cdot x_2} \, Z^{1/2}(\boldsymbol{k}_2)   \,,
\end{align}
where the multiplicity $2!$ mentioned above has been neglected. 
It should be noted that there is no analogous contribution in the 
$t_2 \rightarrow \infty$ limit since the integrand is cut off by 
$\theta(t_3-t_2)$.  

The integral over $d^3 \boldsymbol{x}_2$ results in 
$(2\pi)^3 \delta^3(\boldsymbol{k}_2 - \boldsymbol{p}_2)$ which, when combined 
with $\delta^3(\boldsymbol{k}_1 - \boldsymbol{p}_1 )$, selects a unique 
two-particle in-state 
\begin{multline}
\widetilde{F}^{J}_{\boldsymbol{p}_{m+n}\boldsymbol{p}_1}(q_{r+1},\dots,q_2) =
\int\prod_{j=3}^{r+1} \left\{ d^4x_j\,{\rm e}^{-iq_j\cdot x_j} \right\}
	\, \theta(t_{r+1}-t_r) \cdots \theta(t_4-t_3) \,  
	\\ \times 
	\frac{Z^{1/2}(\boldsymbol{p}_2)}{2E(\boldsymbol{p}_2)} 
	\langle \boldsymbol{p}_{m+n} | \hat{\phi}(x_{r+1}) \cdots \hat{\phi}(x_{m+1}) 
	\, \hat{J}(0) \hat{\phi}(x_m) \cdots \hat{\phi}(x_3) | \boldsymbol{p}_1
	\boldsymbol{p}_2 \rangle_{\rm in,c}\int_{-\infty}^{-t_{\rm c}}dt_2 \,{\rm e}^{-i(q_2^0 - E(\boldsymbol{p}_2)  + i \epsilon)t_2} + \cdots \,. 
\end{multline}
 where we have made the $i \epsilon$-prescription explicit in the $t_2$-dependence. The $t_2$-integral results in the expected pole factor
\begin{gather}\label{e:pole}
	\int_{-\infty}^{-t_{\rm c}}dt_2\, {\rm e}^{-i(q_2^0 - E(\boldsymbol{p}_2)+
	i\epsilon)t_2} = \frac{i}{q_2^0 -E(\boldsymbol{p}_2) + i\epsilon} + \cdots   \,,
\end{gather}
where the ellipsis denotes subleading terms in the expansion of ${\rm e}^{i(q_2^0-E(\boldsymbol{p}_2) + i\epsilon)t_{\rm c}} = 1 + \mathcal O\big [(q_2^0-E(\boldsymbol{p}_2) + i\epsilon) \big ]$.     

The field $\hat{\phi}(x_2)$ has thus been reduced into the right endcap 
isolating the two-particle in-state and corresponding pole factor from 
Eq.~\ref{e:pole}. Similarly, $\hat{\phi}(x_{r+1})$ may be reduced 
into the left endcap by inserting a complete set of two-particle out-states and 
examining the region $t_{r+1}>t_{\rm c} > t_r$. This process is subsequently applied to the left-most and right-most fields to reduce them into the left and right endcaps (respectively), achieving the desired result of  Eq.~\ref{e:tlsz}.


\bibliography{latticen}   
\end{document}